\documentclass{aa}  
\usepackage{graphicx}
\def\fz{photometric redshifts }

\def\lya{Ly$\alpha$ }

\newcommand{\di}{{\it detection}}
\newcommand{\mi}{{\it measure}}
\newcommand{\dii}{{\it detection} image}
\newcommand{\mii}{{\it measure} image}
\newcommand{\cp}{ConvPhot}
\begin{document}
   \title{The GOODS-MUSIC sample: a multicolour catalog of near-IR selected
galaxies in the GOODS-South field
\thanks{Tables~\ref{goodscatA}, \ref{goodscatB}, and \ref{goodscatC} are also
available in electronic form
at the CDS via anonymous ftp to cdsarc.u-strasbg.fr (130.79.128.5)
or via http://cdsweb.u-strasbg.fr/cgi-bin/qcat?J/A+A/}
}

   \author{A. Grazian  \inst{1}
          \and
          A. Fontana  \inst{1}
          \and
          C. De Santis \inst{1}
          \and
          M. Nonino \inst{2}
          \and
          S. Salimbeni\inst{1}
          \and
          E. Giallongo\inst{1}
          \and
          S. Cristiani\inst{2}
          \and
          S. Gallozzi \inst{1}
          \and
          E. Vanzella\inst{2}
}

   \offprints{A. Fontana, \email{fontana@mporzio.astro.it}}

\institute{INAF - Osservatorio Astronomico di Roma, Via Frascati 33,
I--00040, Monteporzio, Italy
\and INAF - Osservatorio Astronomico di Trieste, Via G.B. Tiepolo 11,
I--34131, Trieste, Italy}

   \date{Received August 3, 2005; Accepted November 17, 2005}

   \titlerunning{GOODS-MUSIC multicolour catalog}
 
  \abstract
{}
{
We present a high quality multiwavelength (from 0.3 to 8.0 $\mu$m) catalog
of the large and deep area in the GOODS Southern Field covered by the
deep near--IR observations obtained with the ESO VLT.
}
{
The catalog is
entirely based on public data: in our analysis, we have included the
$F435W$, $F606W$, $F775W$ and $F850LP$ ACS images, the $JHKs$ VLT
data, the Spitzer data provided by IRAC instrument (3.6, 4.5, 5.8 and
8.0 $\mu m$), and publicly available U--band data from the 2.2ESO
and VLT-VIMOS. We describe in detail the procedures adopted to
obtain this multiwavelength catalog. In particular, we developed a
specific software for the accurate ``PSF--matching'' of space and
ground-based images of different resolution and depth
(ConvPhot), of which we analyse performances and limitations. We
have included both $z$--selected, as well as $Ks$--selected objects,
yielding a unique, self--consistent catalog.
The largest fraction of the sample is 90\% complete
at $z\simeq 26$ or $Ks\simeq 23.8$ (AB scale). Finally, we
cross-correlated our data with all the spectroscopic catalogs
available to date, assigning a spectroscopic redshift to more than
1000 sources.
}
{
The final catalog is made up of 14847 objects, at
least 72 of which are known stars, 68 are AGNs, and 928 galaxies with
spectroscopic redshift (668 galaxies with reliable redshift
determination). We applied our photometric redshift code to this data set,
and the comparison with the spectroscopic sample shows that
the quality of the resulting photometric redshifts is excellent, with
an average scatter of only 0.06.  The full catalog, which we named
GOODS-MUSIC (MUltiwavelength Southern Infrared Catalog), including the
spectroscopic information, is made publicly available, together with
the software specifically designed to this end.
}
{}

   \keywords{Galaxies: distances and redshifts - Galaxies: evolution -
Galaxies: high redshift - Galaxies: photometry -
Methods: data analysis - Techniques: image processing}

   \maketitle
%

\section{Introduction}

The advent of the Hubble Space Telescope and of the modern giant
telescopes has opened a new era in observational cosmology, where
galaxy evolution can be traced back to its early stages.  In this
context, deep multicolour imaging surveys are established as a powerful
tool to access the population of faint galaxies with relatively high
efficiency.  An already classical technique (\cite{steidel95},
\cite{madau98}) is used to select high redshift galaxy candidates with
sharp colour selection criteria, which has been shown to be particularly
efficient at $z\simeq 3-4$ (\cite{steidel99}).

Other strategies have focused on surveys that sample the whole
spectral range from the U to the K band, enabling galaxy evolution to be
followed on a wider range of redshifts, and that have triggered the
development and use of \fz to extract information about galaxies at
any redshift. The Hubble Deep Field North (HDFN), and the subsequent Hubble
Deep Field South (HDFS), are the most successful examples of this approach,
which has stimulated an impressively large number of scientific
investigations (see \cite{ferguson00} for a review).

The Great Observatories Origins Deep Survey (GOODS, \cite{dickinson01})
represents the next generation of this kind of survey. It is
based on observations of two separated fields of 10$\times$15 arcmin
each, centred on the HDFN and Chandra Deep Field South (CDFS
hereafter), respectively.  Over a total area that is about 50 times
the combined HDFN/S, it is the result of an impressive effort with the
most advanced facilities.

The available dataset includes ultradeep images from ACS on HST, from
the mid--IR satellite Spitzer, from the ultraviolet satellite GALEX,
from the X-ray satellites Chandra and
XMM, as well as from a long number of ground--based facilities (see
\cite{giavalisco04} for a more detailed presentation).
Spectroscopic follow--up is also available or ongoing from a list
of surveys and facilities (\cite{cimatti02,cowie04,vanzella05}).

In particular, the GOODS southern pointing, located over the CDFS
(centred on $\alpha=03:32:30$ $\delta=-27:48:30$, \cite{giacconi})
and encompassing the ACS Ultra Deep Field (UDF hereafter, \cite{udf})
and the
GMASS\footnote{\sl http://www.arcetri.astro.it/$\sim$cimatti/gmass/gmass.html}
ESO Large Program (Galaxy Mass Assembly ultradeep Spectroscopic Survey,
PI Cimatti),
has been
the target of extensive observations with ESO telescopes, carried on
in the spirit of public surveys. To date, a major effort has been
devoted to deep imaging observations in the $J$, $H$, and $Ks$ bands
with the VLT-ISAAC infrared imager. The final area covered by the $Ks$
images approaches 140 arcmin$^2$ to a $1\sigma$ limiting depth of
typically 26-27 mag/arcsec$^2$ (AB scale), making it a unique
combination of depth and size in the near--IR. Additional observations
in the U band have also been obtained, and a large spectroscopic 
follow-up is currently underway.

This paper describes the multicolour catalog, which we named
GOODS-MUSIC\footnote{Note that our
sample is definitely unrelated to the data and activities of the
MUSYC collaboration, {\tt http://www.astro.yale.edu/musyc/}}
(MUltiwavelength Southern Infrared Catalog), that we obtained
using the public imaging data available in the GOODS--CDFS region.  It
includes two $U$ images obtained with the ESO 2.2m telescope and one
$U$ band image from VLT-VIMOS, the ACS--HST images in four bands, the
VLT-ISAAC $J$, $H$, and $Ks$ bands as well as the Spitzer images in at
3.5, 4.5, 5.8, and 8 $\mu m$. Most of these images have been made
publicly available in the coadded version by the GOODS team, while the
$U$ band data were retrieved in raw format and reduced by our
team.

From this dataset we obtained both a $z$--selected, as
well as a $Ks$--selected, sample. We also collected all the available
spectroscopic information and cross-correlated the spectroscopic
redshifts with our photometric catalog. For the 
unobserved fraction of the objects, we  applied our \fz code to
obtain well-calibrated photometric redshifts, which are described in a
forthcoming paper (Vanzella et al. in preparation), together with the
redshift based on the neural network technique (\cite{vanzellaNNz}).

The resulting dataset will be used for several types of scientific output, for
which we refer to forthcoming papers. At the same time, we make the
dataset publicly available, along with some software specifically
developed for this project, all of which can be of wider interest, and which
we first describe here.

The paper is organised as follows. In Sect. 2, we briefly summarise
the available dataset. In Sect. 3, we discuss the procedure for object
detection in the ACS-$z$ image, and the corresponding measurement of total
magnitude. In Sect. 4, we discuss the procedure adopted to estimate
colours of $z$--selected objects, which has requiered the development
of a new software code, which we explain in some detail. In Sect. 5, we
focus on the $Ks$--selected sample, describing how we identify the
objects that are detected in $Ks$ and not in $z$, and how we measure
their total magnitude and colours. In Sect. 6 we briefly list the
spectroscopic observations that we used to assign spectroscopic
redshifts, while Sect. 7 discusses the photometric redshift technique applied.
Section 8 finally describes the format and
the technical details adopted for the public release.

All magnitudes are given in AB scale, and a $\Lambda$-dominated
cosmology ($\Omega_m =0.3$, $\Omega_\Lambda =0.7$ and $H_0 = 70$ km
s$^{-1}$Mpc$^{-1}$) has been adopted throughout the paper.

The data and photometric redshift catalogs are also available on
line at the WEB site {\sf http://lbc.oa-roma.inaf.it/goods}.


\section{Imaging data}

\subsection{GOODS-south ACS data}

In this paper we will focus on the southern field of the GOODS survey,
located in the CDFS.

Images are made available for all 4 GOODS filters, which are those
used in the UDF observations: F435W, F606W, F775W, and F850LP.
In the following, we will refer to these images as $B$, $V$, $i$, and
$z$, respectively.  Of this data set, in this work we use the data
release 1.0 provided by the GOODS team (\cite{giavalisco04}).  Table
\ref{summaryGOODS} summarises the characteristic of the each pointing of
the GOODS ACS data.

The V1.0 ACS GOODS images released by the GOODS team were divided into a
grid of 4 by 5 8K x 8K frames for each band. Since we were only interested
in the portion of the image covered by the J and Ks
observations, we considered only the tiles that had an overlap (at least
partial) with the IR images. Of these images, we built a
single mosaic for each band (the software used for this purpose, named
$fitstile$, is available on our web site): each resulting frame is
3.4Gb wide, which can still be handled by SExtractor in a single
run, provided it is compiled with the ``large file'' option.  Although
this solution is somewhat impractical (for instance, common
visualisation tools fail to display such large images), it allows us to
avoid the complications arising when catalogs are obtained on
overlapping images, such as inhomogeneities in the large scale
background, different object detection, and other border effects.

\subsection{The ISAAC data}

The GOODS field of the CDFS was the target of
a deep imaging campaign in the near infrared with the ESO telescopes.
A large field (20 by 20 arcmin) was covered with SOFI at a shallow
magnitude limits in the J, H, and Ks bands. The GOODS-CDFS is
being covered by much deeper observations in the same NIR bands with the
ISAAC instrument. These  data have been partially released by ESO and
will be used in this work. A full documentation of this data set is
presented in Vandame et al. (in preparation), while details of the on-going
GOODS program at ESO are given at {\sl
http://www.eso.org/science/goods/products.html}.

The version 1.0 data release, analysed with the ESO-MVM pipeline for imaging
data reduction (\cite{vandame02}), includes
22 fully reduced VLT/ISAAC fields in J and Ks bands covering about 140
$arcmin^2$ of the GOODS region in the CDFS.

\begin{figure}
\includegraphics[width=9cm]{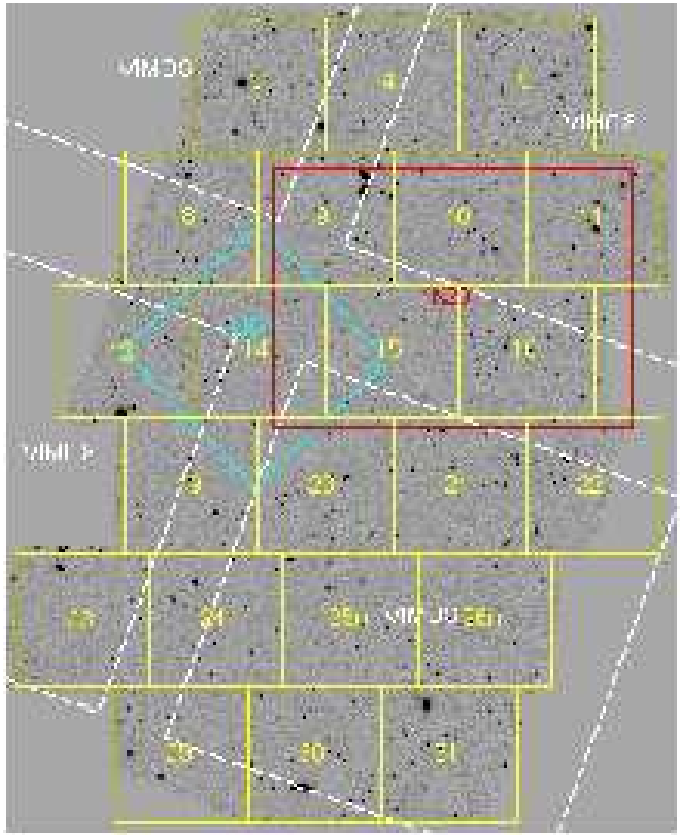}
\caption{The ACS GOODS-S field with superimposed the ISAAC tiling
made by ESO to cover the whole field in J and Ks. The field shown is the
ACS in the z-band limited to the area common to ACS and ISAAC
observations.
The cyan square
marks the position of the ACS UDF. The NICMOS UDF Treasury observations
cover a field that is inside the UDF. The red rectangle marks the position
of the K20 survey. The four white quadrants show the actual coverage of
VLT-VIMOS U band imaging; large gaps are visible, since the observing program
has not yet been completed.}
\label{newrelease}
\end{figure}

Figure \ref{newrelease} shows the position of the released ISAAC fields
(yellow) superimposed on the optical image of the CDFS by ACS.  The
cyan square shows the position of the UDF on the GOODS-CDFS field,
while the red rectangle shows the layout of the K20 survey. The four white
quadrants show the actual coverage of the VIMOS U band imaging.
The H band observations cover
only half of the entire survey, and are only limited to twelve fields.
They have not been released by the present V1.0 data release but were part
of a previous (V0.5) EIS data release and are public on the ESO
archive.

The typical exposure times range from 3 to 6 hours, the seeing
ranges from 0.4 to 0.6 arcsec, and it is typically below 0.5 arcsec.
We measured the value of the seeing for each ISAAC field by
averaging all the bright stellar objects from the image, after
normalisation of their profiles to unit flux. For each ISAAC field, we
also obtained a ``convolution kernel'' ${\cal K}$, which is the
transformation matrix that converts the ACS PSF to the relevant ISAAC
one that we shall use for the colour estimate. This transformation was
computed in Fourier space, as ${\cal K}'=PSF'_{ISAAC} / PSF'_{ACS}$,
where the prime refers to the Fourier transform. An optimal Wiener
filter, which suppresses the high frequency fluctuations,
was applied in the Fourier domain to remove the effects of
noise.

%

\subsection{U-band data: 2.2-WFI and VLT-VIMOS}

We searched the ESO public archives for deep imaging with the aim of
complementing the spectral coverage by the GOODS
team. Particularly important is the U band, which improves the
photometric redshift estimate, 
in particular for the lowest redshift ($z\le 0.5$), when the U
band still probes the blue side of the 4000 \AA~ break
and, most important, is fundamental for identifying $z\sim 3$
galaxies (U-dropouts).

In the ESO Science Archive there are U-band images taken with the wide
field imager (WFI) at La Silla (Chile), which are part of the EIS
public survey (\cite{arnouts01}), as well as recent images with the
VLT-VIMOS imager. 

The WFI images were obtained in two filters, the so-called $U_{35}$
and $U_{38}$. The $U_{38}$ is the standard Bessel U with a peak efficiency of
50\%, centred around $\lambda_c=3800$ \AA, while the $U_{35}$ is a bit
bluer, $\lambda_c=3500$ \AA, with higher efficiency ($\sim
80\%$), but with a red leak at $\lambda\ge 8000$ \AA, where the WFI
CCD is still sensitive. The net exposure time is 53654 s ($\sim 15 h$)
and 75100 s ($\sim 21 h$) for the $U_{35}$ and $U_{38}$ filters, respectively.
The U band image of VIMOS is based on a redder filter ($\lambda_c=3900$ \AA)
and has an exposure time of 10000 seconds in total.
The coverage of the GOODS-CDFS field, however, is partial, since the
observing program has not yet been completed.

We decided to use these images ($U_{35}$) even if they are affected by red
leakage: by inserting the correct (with red leak) filter transmission
curve to the photometric redshift procedure one is able to mimic the
behaviour of distant galaxies in the $U_{35}$ filter.
The VIMOS images available on Dec 2004 cover a significant
fraction of the field (about 60\%), although they exhibit the large
gaps arising from the array disposition (see Fig. \ref{newrelease}).

The raw U-band images were collected from the ESO Science Archive and reduced
using Python scripts and the IRAF task {\sl mscred} following the
instruction of
\cite{valdes02}. The total exposure times of these images surpass
the previous data release of EIS-DPS, which only reached 43200 and
61200 seconds in the $U_{35}$ and $U_{38}$, respectively. The image quality
of these
images is good, with a seeing of 0.9-1.1 arcsec and a magnitude limit of
26.5-27.0 (AB mag) at 5 sigma in a photometric aperture of 3 arcsec.
The U image of VIMOS had good seeing conditions (0.7-0.8 arcsec) and reached
a deeper magnitude limit (28 AB, 5$\sigma$).

\subsection{IRAC data from the Spitzer Legacy Program}

The GOODS survey also incorporates a Spitzer Space Telescope Legacy
Program to carry out the deepest observations with this facility at
3.6 to 24 microns, to study galaxy formation and evolution over a wide
range of redshift and cosmic lookback time.

The first and second Spitzer data releases (DR1 and DR2) consist of
``best-effort'' reductions of data taken with the infrared array camera
(IRAC, \cite{fazio04}) on-board Spitzer.  These are images from the
two epochs of the ``superdeep'' IRAC observations for each of the two
GOODS fields (Dickinson et al., in preparation).

These fields were imaged at $3.6-8\mu$ with IRAC, with a
mean exposure time per position of approximately 23 hours per
band (doubled in the overlap strip containing the UDF),
reaching far deeper flux limits than observations planned for the
Guaranteed Time programs.

The IRAC observes simultaneously in all four channels, with channels 1 and
3 (3.6 and 5.8 microns) covering one pointing on the sky, and channels
2 and 4 (4.5 and 8.0 microns) covering another. The two IRAC
fields of view are separated by about 6.7 arcmin in the focal plane,
and the long axes of the GOODS fields are oriented along the direction
separating the two IRAC fields of view. The consequence of the 2x2
mapping pattern is that, in a given observing epoch, the area covered
by channels 1+3 and the one covered by channels 2+4 have a small region
of overlap (about 3 arcmin). The GOODS-S four-channel overlap region
includes the Hubble UDF (\cite{udf}).

\begin{figure}
\includegraphics[width=9cm]{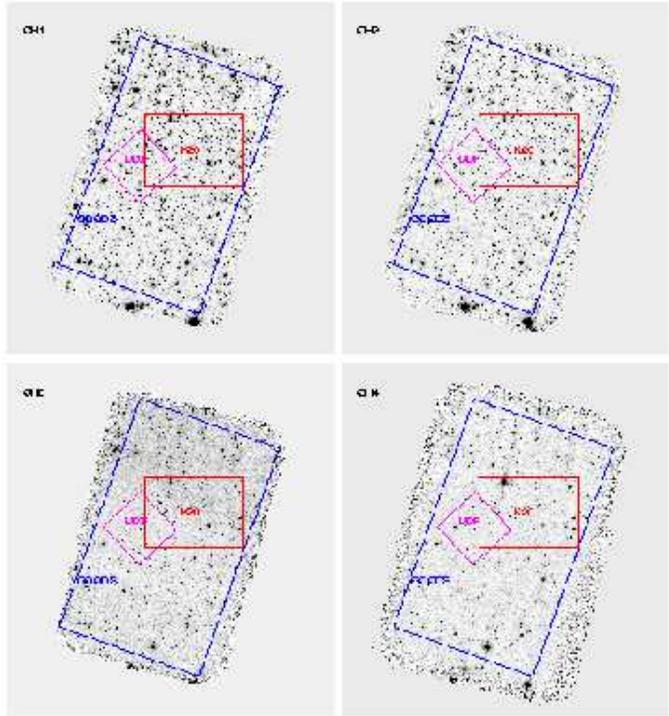}
\caption{The four channels of the IRAC images for the GOODS-South
field. The pointings are chosen to have the UDF in the overlapping regions,
where the exposure time is twice that in the external parts. The K20 layout
is shown, as well.
}
\label{irac}
\end{figure}

The first and second epochs of IRAC images publicly released by the
GOODS Team, cover the entire GOODS-CDFS field at 3.6, 4.5, 5.8 and 8.0
$\mu m$ (see Fig. \ref{irac}).
Table \ref{summaryGOODS} gives detailed information for the GOODS-CDFS
survey used in this paper, with the wavelengths, area covered, and magnitude
limits for all the filters available at this moment.

\begin{table*}
\caption[]{A summary of the photometric data of the GOODS-South field
used in this work}
\begin{tabular}{p{2cm}cccccccc}
\hline
\hline
FILTER & $\lambda_c$ & $\Delta \lambda$ & EXPTIME & FWHM$^a$ & PIXSCALE & ZP
& AREA & MAGLIM$^b$ \\
 & \AA & \AA & s & arcsec & arcsec/px & AB & $arcmin^2$ & 90$\%$ \\
\hline
$U_{35}$     &  3590 &   222 & 53654 & 0.90 & 0.23 & 28.520   & 143.2 & 25.5 \\
$U_{38}$     &  3680 &   170 & 75100 & 1.10 & 0.23 & 28.755   & 143.2 & 24.5 \\
$U_{VIMOS}$  &  3780 &   197 & 10000 & 0.80 & 0.20 & 32.500   &  90.2 & 26.5 \\
B (F435W)    &  4330 &   508 &  7200 & 0.12 & 0.03 & 25.65288 & 143.2 & 27.5 \\
V (F606W)    &  5940 &  1168 &  6000 & 0.12 & 0.03 & 26.49341 & 143.2 & 27.5 \\
i (F775W)    &  7710 &   710 &  6000 & 0.12 & 0.03 & 25.64053 & 143.2 & 26.5 \\
z (F850LP)   &  8860 &   554 & 12000 & 0.12 & 0.03 & 24.84315 & 143.2 & 26.0 \\
$J_{ISAAC}$  & 12550 &  1499 & 12600$^c$ & 0.45$^c$ & 0.15 & 26.000 & 143.2 & 24.5 \\
$H_{ISAAC}$  & 16560 &  1479 & 18000$^c$ & 0.45$^c$ & 0.15 & 26.000 &  78.0 & 24.3 \\
$Ks_{ISAAC}$ & 21630 &  1383 & 23400$^c$ & 0.45$^c$ & 0.15 & 26.000 & 143.2 & 23.8 \\
$CH1_{IRAC}$ & 35620 &  3797 & 82800 & 1.60 & 0.60 & 22.416 & 143.2 & 24.0 \\
$CH2_{IRAC}$ & 45120 &  5043 & 82800 & 1.70 & 0.60 & 22.195 & 143.2 & 23.4 \\
$CH3_{IRAC}$ & 56860 &  6846 & 82800 & 1.90 & 0.60 & 20.603 & 143.2 & 22.0 \\
$CH4_{IRAC}$ & 79360 & 14797 & 82800 & 2.00 & 0.60 & 21.781 & 143.2 & 22.0 \\
\hline
\end{tabular}
\label{summaryGOODS}
\begin{list}{}{}
\item a)
For ground-based images, $FWHM$ corresponds to the seeing, while for
space-based data it is the PSF of the instrument.
\item b)
The limiting magnitudes (at 90 \% completeness)
are the mean values on the field, averaged on the positions and areas
of every object.
\item c)
For J, H, and Ks filters, we report only the mean value of seeing and exposure
time: detailed values can be found in Vandame et al. (in preparation).
\end{list}
\end{table*}

\subsection{Accurate estimation of photometric errors}

The correct evaluation of the photometric errors for astronomical
sources is bound to the possibility of having a realistic variance image
associated with the measure image. It has become a common feature of
the data reduction pipelines to produce ``weight map'' images that
are usually proportional to the exposure time of each pixel and,
therefore, provide a measure of the relative S/N across the image.
These images can be used by SExtractor to obtain a first map of the
background r.m.s., which is the main source of noise for faint sources.
However, the interpolations introduced by drizzling the images
(shifting, rotating, correcting distortion, and subsampling pixels
onto a finer grid) result in correlations between pixels in the
drizzled science images.  Therefore, the apparent r.m.s. background noise
that one measures in the image is smaller than the actual r.m.s., due to
the effects of these correlations (for a more detailed discussion of
weight map conventions and noise correlation in drizzling, please see
\cite{casertano00}, especially Sect. 3.5 and Appendix A).

To overcome this problem, we used a simple and efficient method
to derive the r.m.s. from the scientific image itself, based on the
correlation matrix of the pixels of the background. Indeed, as we explain
better in Appendix A, the true r.m.s. of an image is given by:
\begin{equation}
\sigma_{true}^2=\sum_{j=-m}^{+m}\frac{1}{N} \sum_{i=1}^{N} y_i y_{i+j} \ ,
\end{equation}
where N is the number of pixels used for the r.m.s. calculation, $y_i$ the
value of the $i$-th pixel on the source-subtracted image, and $m$ the
correlation length of the noise, typically of 10-20 pixels.
 
In practice, we used SExtractor to convert the
weight maps (available for each image of our data set) into
the corresponding image of apparent r.m.s. We then estimated the
correlation matrix in a homogeneous section of the image to compute
the accurate r.m.s., and finally normalised the apparent
r.m.s. image accordingly.

We verified this procedure in the ACS images, where the weight images
(\cite{giavalisco04})
coincide with the expected inverse variance maps (i.e., $1/\sigma^2$) per
pixel.  In this case, we obtained an absolute r.m.s. of 0.001396,
comparable to the value 0.0014 obtained with a traditional
approach\footnote{We polluted the $z$ band images of GOODS with
synthetic stars of different magnitudes and derived the true magnitude
errors by the variance over 100 realisations}, as well as with the
publicly available weight maps (i.e., $1/\sigma^2$).

For the U, ISAAC, and Spitzer images, we applied the same technique
and found that the true r.m.s. is typically a factor 1.4 times
higher than the r.m.s. that was directly measured on the science frames.
As explained above,
in the case of the ACS images, we used the available weight maps to
compute the r.m.s. image by simply inverting the square root of the weight
map.

\subsection{Magnitude limits and effective areas}

The total area of the survey is the intersection of the images in B,
V, $i$, $z$, J, and Ks bands. The $U_{35}$ and $U_{38}$ images coves a
much larger field of view (FoV), approximately 30 by 30 arcmin,
the IRAC images completely cover the field layout,
while the VIMOS-U and the H band cover about half of the total
field.  Images in V, $i$, and $z$ cover the same area, while B is smaller
due to the lower number of dithering steps. Since we are interested in
$z$ and $Ks$--selected samples with good photometric redshift accuracy,
we limit the sample to the area given by the intersection of the $B$
and $Ks$ images.  In this sample, all the objects have
$U_{35}U_{38}BVizJKs$ and IRAC coverage, and about a half also have VIMOS-U and
$H$ band observations.  The total area is 143.2 sq. arcmin. and the layout
of the resulting field is shown in Fig. \ref{newrelease}.

\begin{figure}
\includegraphics[width=9cm]{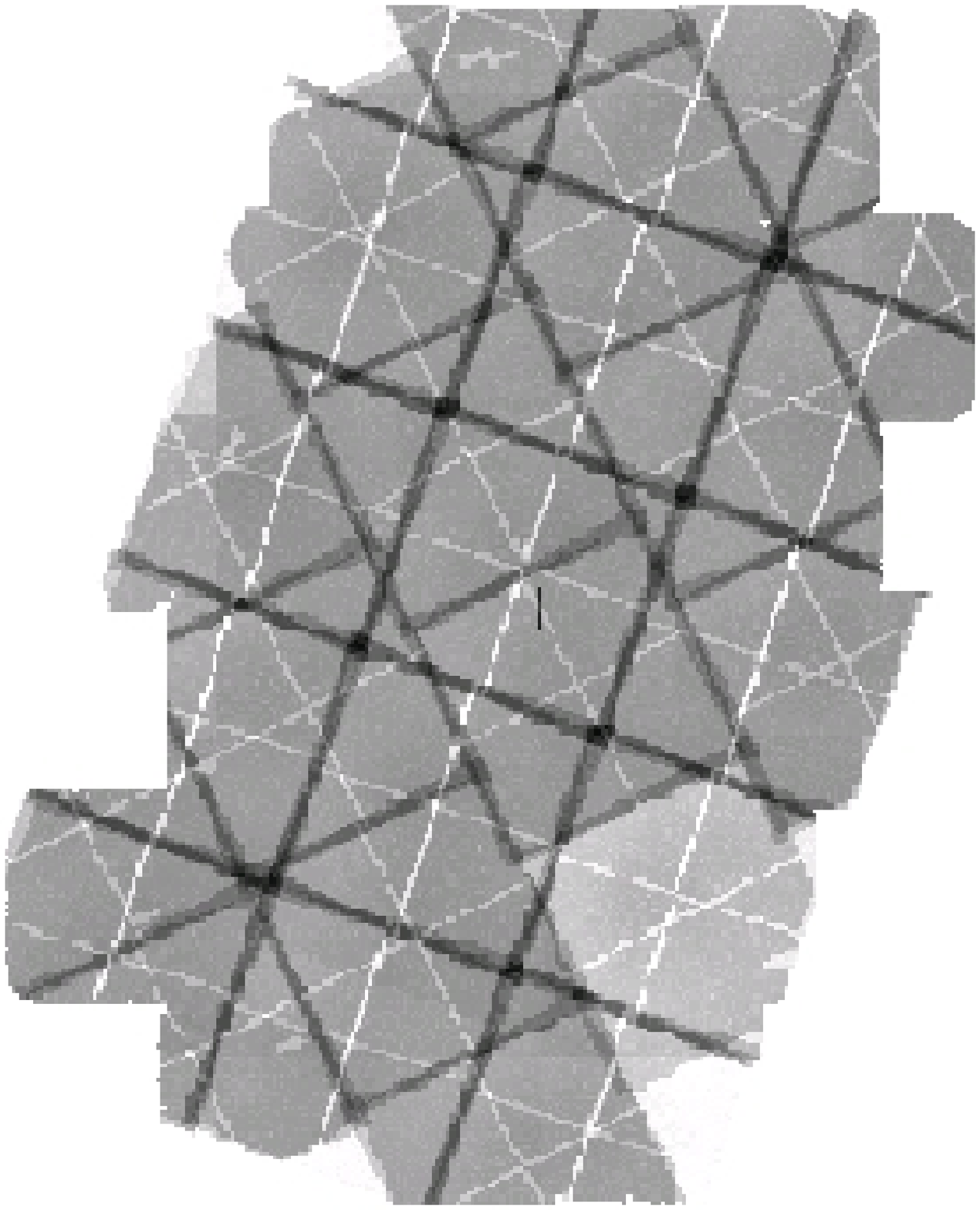}
\includegraphics[width=9cm]{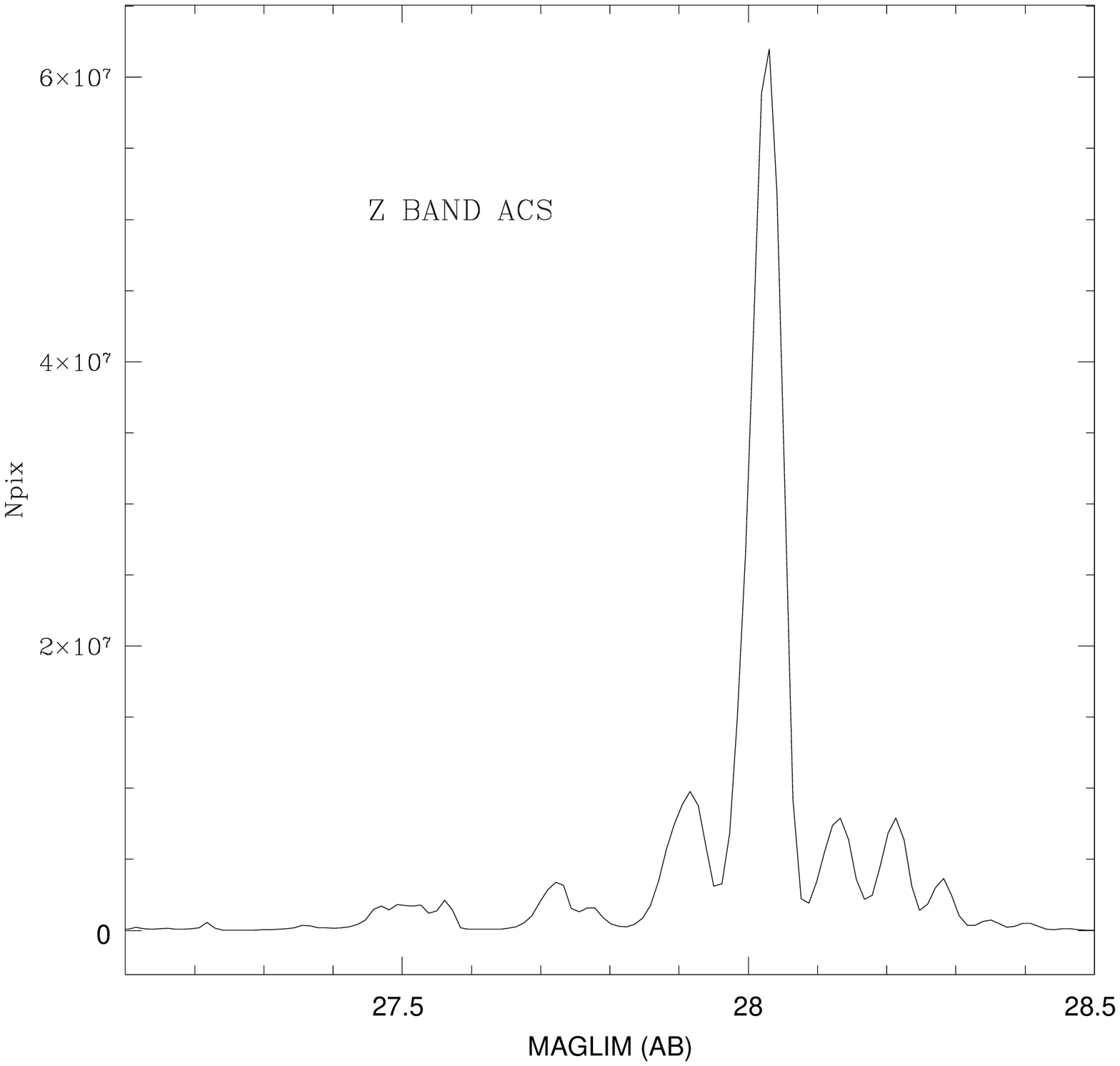}
\caption{The magnitude limits (at 1 sigma and in 1 sq. arcsec) for
the $z$ band in the GOODS area. Dark areas correspond to deeper
exposures, while white zones are much shallower ones, as also shown in
the histogram of the magnitude limits. The geometry of the dark and white
areas depends on the dithering strategy adopted.}
\label{fieldZ}
\end{figure}

\begin{figure}
\includegraphics[width=9cm]{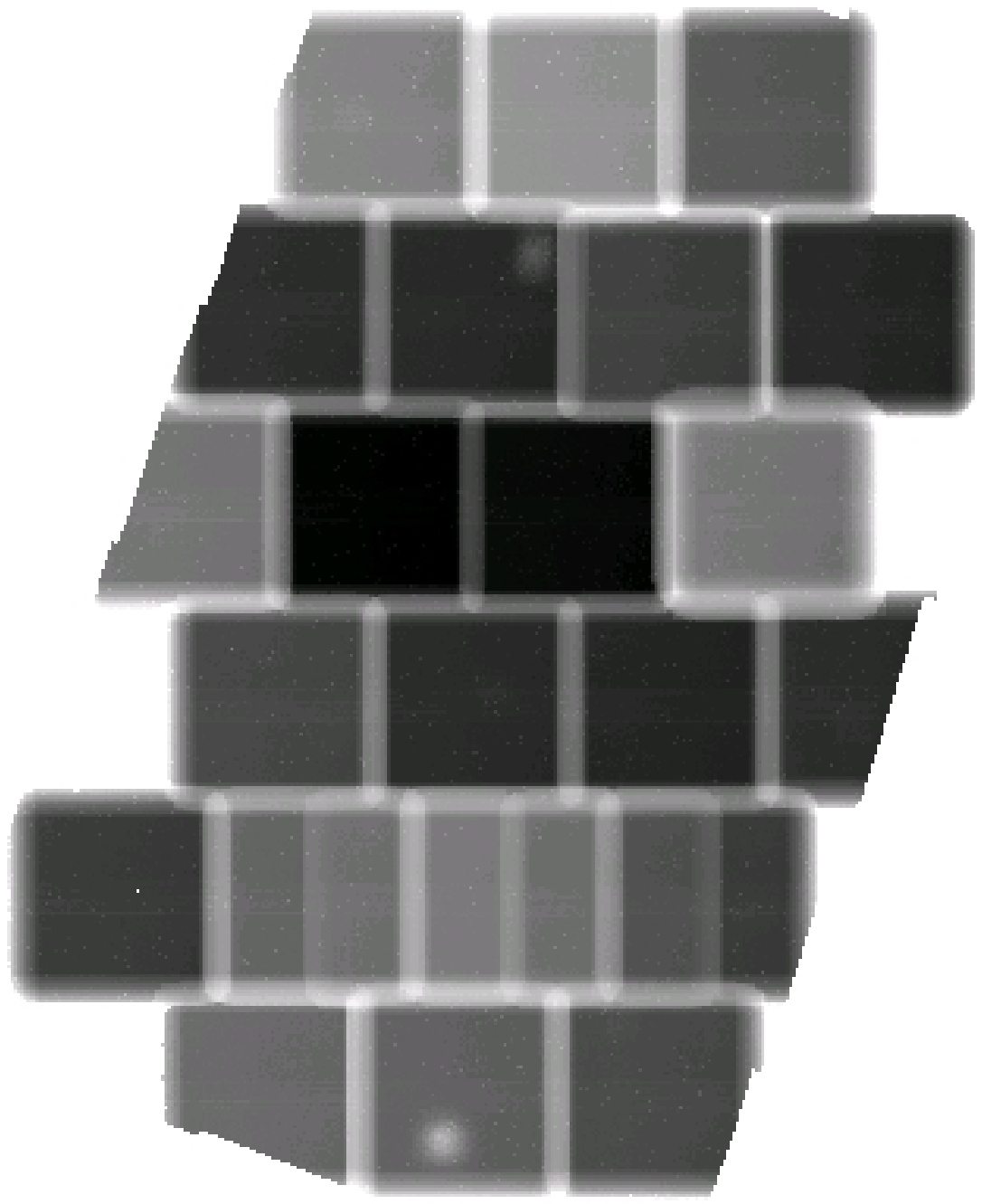}
\includegraphics[width=9cm]{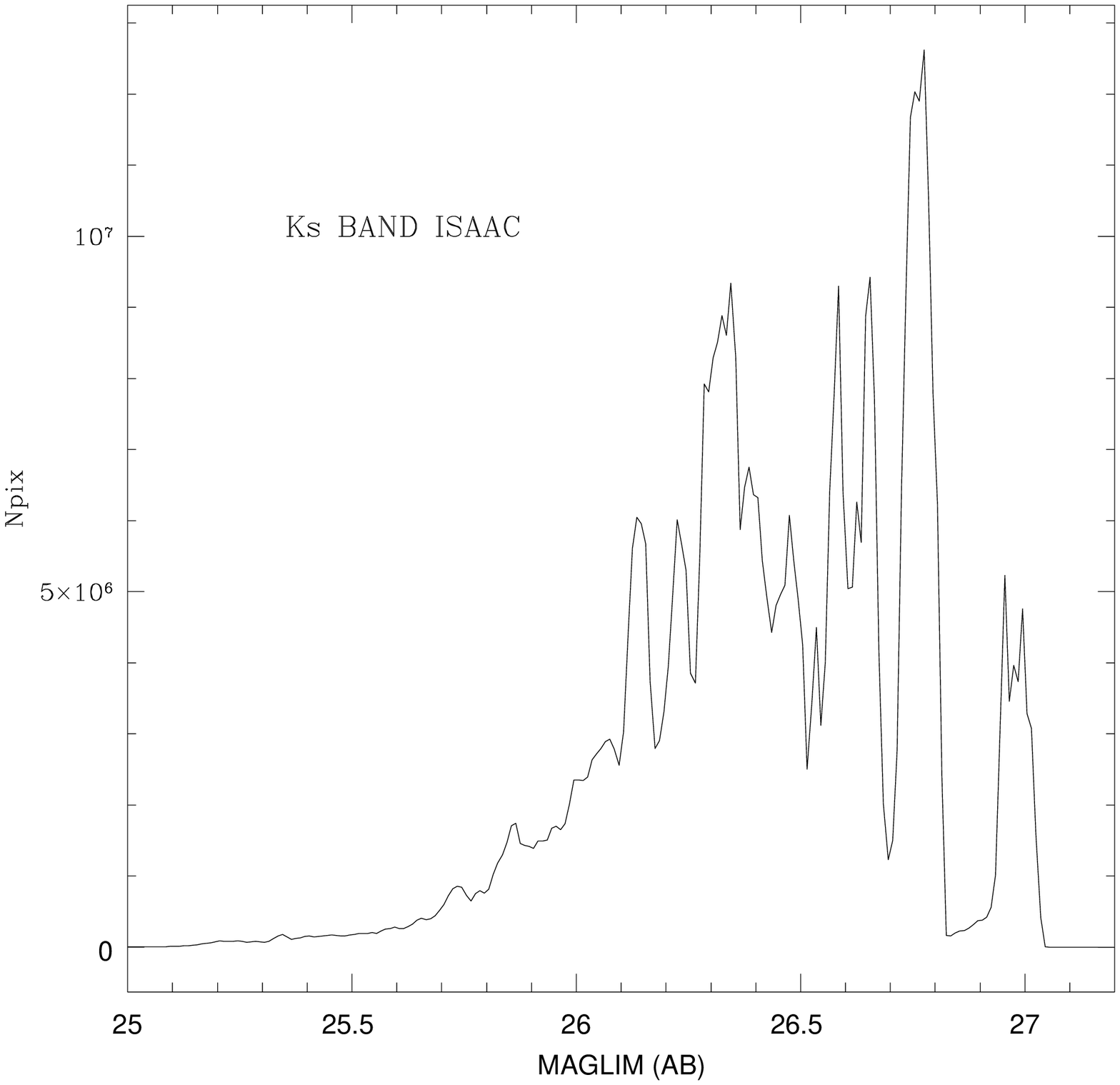}
\caption{The magnitude limits (at 1 sigma and in 1 sq. arcsec) for
the Ks band in the GOODS area. Dark areas correspond to deeper exposures,
while white zones are much shallower ones, as also shown in the histogram
of the magnitude limits.}
\label{fieldKs}
\end{figure}

Associating a limiting magnitude to this area is not trivial, since
the exposure maps in optical and NIR are different and very
complicated.  We first decided to estimate an accurate map of the
magnitude limits as a function of position using r.m.s. images in $z$ and
Ks. The limiting magnitudes are computed at 1 sigma level and in an
area of 1 sq. arcsec. Figures \ref{fieldZ} and \ref{fieldKs} show this
in the GOODS area, as well as the relevant
distribution of the magnitude limit. Dark areas represent deeper
exposures that correspond to the faintest magnitude limits.
In the $z$ band image we identified 6 main areas with a different
interval in the magnitude limits; in Ks band the magnitude limit
intervals identified are 6 (see Table \ref{maglimtab}).

This information was used as follows. We associated the $1-\sigma$
magnitude limit to each
detected object (either in the $z$ or in the $Ks$--selected sample)
as obtained from the ``magnitude limit
maps'' at the corresponding position. The simulations described in
Sect 3. made it possible to translate this $1-\sigma$ magnitude limit into a
completeness limit for the object detection. Each catalog was
therefore separated according to the magnitude limit, using the bins
listed in Table \ref{maglimtab}. As a result, each catalog is actually
the combination of 6 (for $z$ and for $Ks$) catalogs with different
magnitude limits. They should be used as independent
catalogs with different magnitude limits and areas. This solution can
be easily applied to all the cases where volume--sensitive statistics
have to be applied, as in the case of luminosity densities or
luminosity functions. Analyses that are sensitive to the position
information (as clustering) instead require the use of the ``magnitude
limit maps''. In the publicly released catalog we provide the
corresponding magnitude limit both in $z$ as in $Ks$ with each object.

\begin{table}

\caption[]{Magnitude intervals that we used to divide the overall sample into 
subsamples with well-defined magnitude limits and corresponding
areas for the $z$ and $Ks$ bands. See also Figs. \ref{fieldZ} and
\ref{fieldKs}.  $MAGLIM1$ and $MAGLIM2$ (in magnitudes per
sq. arcsec) define the limits of each magnitude bin. $MAGTOT$ is the
corresponding limit for the total magnitude of the objects, obtained
using the simulations described in Sect. 3.2.
}
\begin{tabular}{lcccc}
\hline
\hline
{\small FILTER} & {\small MAGLIM1} & {\small MAGLIM2} & {\small AREA}
& {\small MAGTOT} \\
       & $1\sigma$ & $1\sigma$ & $arcmin^2$ &   \\
\hline
z & 26.65 & 27.65 & 7.16667 & 24.65 \\
z & 27.65 & 27.85 & 6.40742 & 25.65 \\
z & 27.85 & 27.95 & 15.4276 & 25.85 \\
z & 27.95 & 28.10 & 89.9600 & 25.95 \\
z & 28.10 & 28.18 & 10.3641 & 26.10 \\
z & 28.18 & 28.90 & 13.8741 & 26.18 \\
\hline
Ks & 24.70 & 26.08 & 10.7698 & 21.60 \\
Ks & 26.08 & 26.20 & 8.95936 & 22.98 \\
Ks & 26.20 & 26.50 & 40.5733 & 23.10 \\
Ks & 26.50 & 26.70 & 38.2161 & 23.40 \\
Ks & 26.70 & 26.90 & 34.8572 & 23.60 \\
Ks & 26.90 & 27.10 & 9.82067 & 23.80 \\
\hline
\end{tabular}
\label{maglimtab}
\end{table}


\section{Object detection}

\subsection{ACS detection}

The $z$ band of ACS GOODS frames was used as a detection image to
build a catalog for the GOODS-South field. The advantages are twofold:
first, the morphological details and the resolution of the space-based
images cannot be obtained with currently available ground-based
instruments in the optical. Second, the longest wavelength available
in the ACS instrument makes it possible to detect high redshift objects
(i-dropout).  Although
fainter sources in the $i$ band exist that are not or barely detected
in the $z$ band images, the significance of their detection is less
than 90 $\%$ complete, so they cannot be used as a statistically complete
sample and are not considered here.

The optical catalog of GOODS-South field has been produced running
SExtractor (\cite{bertin96}) using the $z$ band big
mosaic as detection image. We used an external flag image to eliminate
the borders of the GOODS field.

Because of the large number of tunable parameters, the use of
SExtractor on a given data set is not straightforward.  After the
first checks on the catalog, we realized that it is impossible to find
a unique set of parameters to obtain an optimal detection over large
FoVs, since in large and deep areas the objects have a wide range of
dimensions, going from the small and faint objects to the large,
low-surface brightness but with high total luminosity near galaxies. In
particular, the most critical parameters are those regulating the
deblending and extension of the object (DETECT-MINAREA,
DETECT-THRESH, DEBLEND-MINCONT, and DEBLEND-NTHRESH).

To provide an optimal detection technique throughout all the GOODS
field, we modified the SExtractor code, following our
experience on the HDFS, described by Vanzella et al. (2001). The idea is to
adopt a general set of SExtractor parameters for the global image and
to define small areas (subframes) where a set of {\em ad hoc}
parameters are adopted to optimise the detection of problematic
objects. In this case we adopted a set of parameters tuned to
detect the faint, compact objects to obtain the bulk of the catalog.
With these parameters, SExtractor typically oversplits bright,
extended, and irregular sources (such as face--on spirals or low-surface
brightness objects) in many fragments. To detect these
objects, we obtained another version of the catalog using an
extreme set of parameters (i.e. requesting very large area and no
deblending) and visually inspected all the objects with magnitude
$z<23$ and all
the objects with spectroscopic redshift. In about 15\% of the
inspected objects we decided to adopt tailored SExtractor parameters
in a corresponding ``subframe''. An investigation at fainter
magnitudes has shown that such cases become very rare at $z>23$, so
we ignored them.

We also slightly modified the SExtractor code to ensure that
the ``back-annulus'' is at least 1 arcsec wide (as suggested by the 
GOODS team) and to correct a bug for the estimation of the isophotal-corrected
magnitude in the ``dual image'' mode when the r.m.s. of detection and measure
image differ by several orders of magnitude (see also the SExtractor
mailing list).

\begin{table}
\caption[]{SExtractor parameters for detection in the $z$ band}
\begin{tabular}{cc}
\hline
\hline
PARAMETER & VALUE \\
\hline
{\small DETECT\_MINAREA} & 13 \\
{\small DETECT\_THRESH} & 0.9707254 \\
{\small ANALYSIS\_THRESH} & 0.9707254 \\
{\small DEBLEND\_NTHRESH} & 32 \\
{\small DEBLEND\_MINCONT} & 0.05 \\
{\small BACK\_SIZE} & 120 \\
{\small BACK\_FILTERSIZE} & 9 \\
{\small BACKPHOTO\_THICK} & 100 \\
{\small CLEAN\_PARAM} & 3 \\
\hline
\end{tabular}
\label{sexpar}
\end{table}

Turning again to the global SExtractor configurations, we adopted
the parameters listed in Table \ref{sexpar}.
The adopted value for ${\small DETECT\_THRESH}$ corresponds to a 3.5$\sigma$
detection over an area equal to 13 pixels.
The high value for the cleaning parameter is due to the need to
have connected pixels in the SEGMENTATION image: with a lower cleaning
parameter it happens that a segmentation with the same identification
number can be disconnected (not contiguous pixels).

This parameter set was derived after detailed simulations and
inspections of the background--subtracted and object--subtracted images, in
order to maximise the completeness and to minimise the number of spurious
objects or artifacts during the detection process. We note in particular
that wide boxes for the background estimations are required to 
prevent large bright sources to distort the background map.

We compared our catalog with version r1.1 of the ACS
multi-band source catalog released by the GOODS Team
(\cite{giavalisco04}).  Excluding objects that are fainter than the detection
limit, there are few galaxies with a difference of 1
magnitude or more between our z band estimation and the one provided by
Giavalisco et al. (2004). These are systematically blended objects where we
used the sub-frames with adaptive parameters to optimise the
detection. Considering only isolated objects, the comparison gives
$<|z-z_{r1.1}|>=0.01$ with $\sigma=0.12$.  This result is not
surprising, since both catalogs use the AUTO magnitudes provided by
SExtractor as an estimate of the total magnitude. We made no attempt
to correct for the known biases of the AUTO (or ``Kron'') magnitudes.

\subsection{Simulating the completeness of the detection}

We used a typical value for the r.m.s. in the $z$ band of 0.0014
(corresponding to a limiting magnitude of $z=28.18$ AB at 1 sigma in 1
sq. arcsec.) to simulate the completeness of our detection criterion
as a function of the magnitude of the sources and their size (half
light radius). Note that $z=28.18$
as a magnitude limit corresponds to the deepest exposures for the $z$
band; for the less exposed areas, the completeness magnitude scales
down to brighter values.

\begin{figure}
\includegraphics[width=9cm]{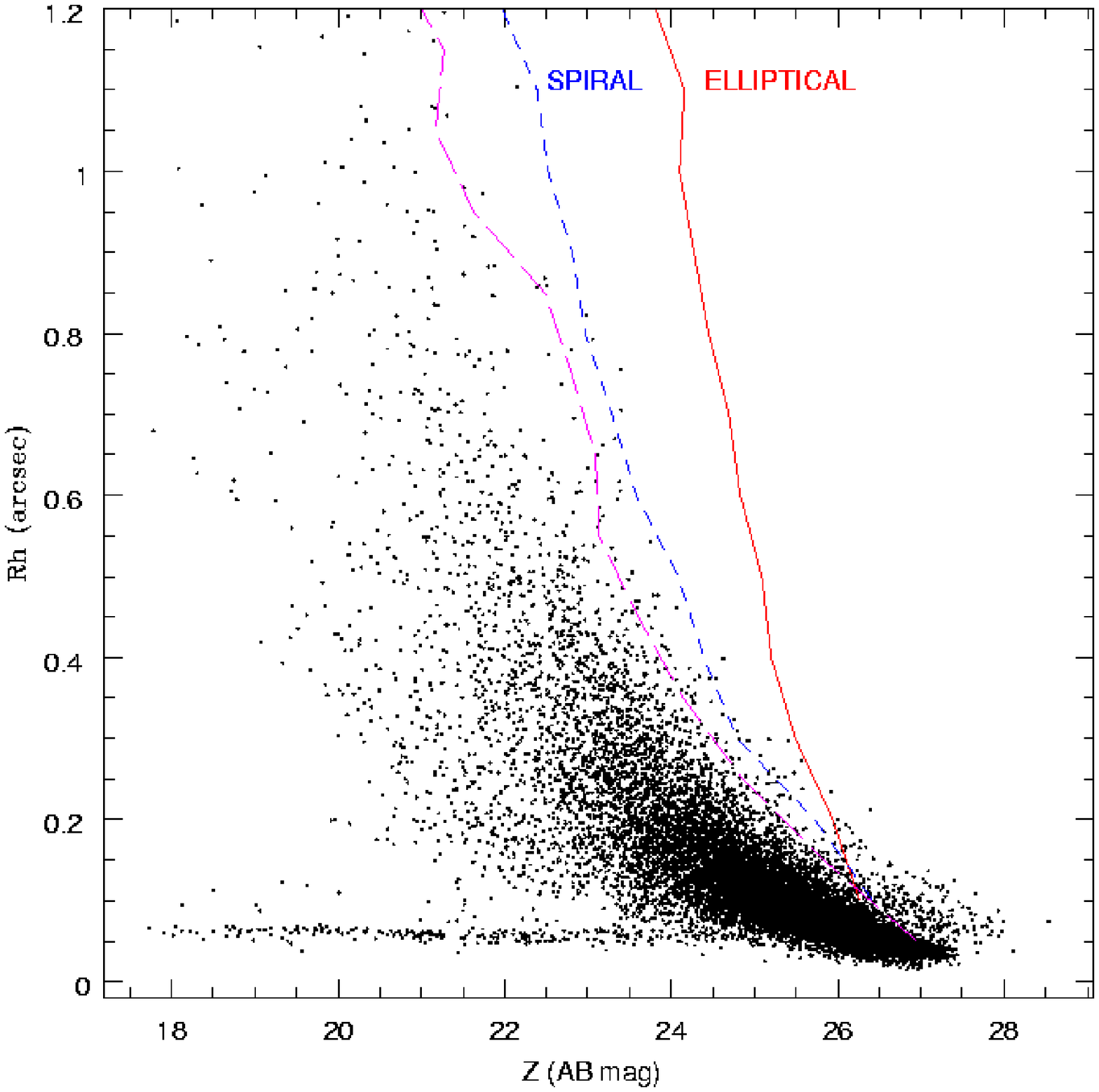}
\includegraphics[width=9cm]{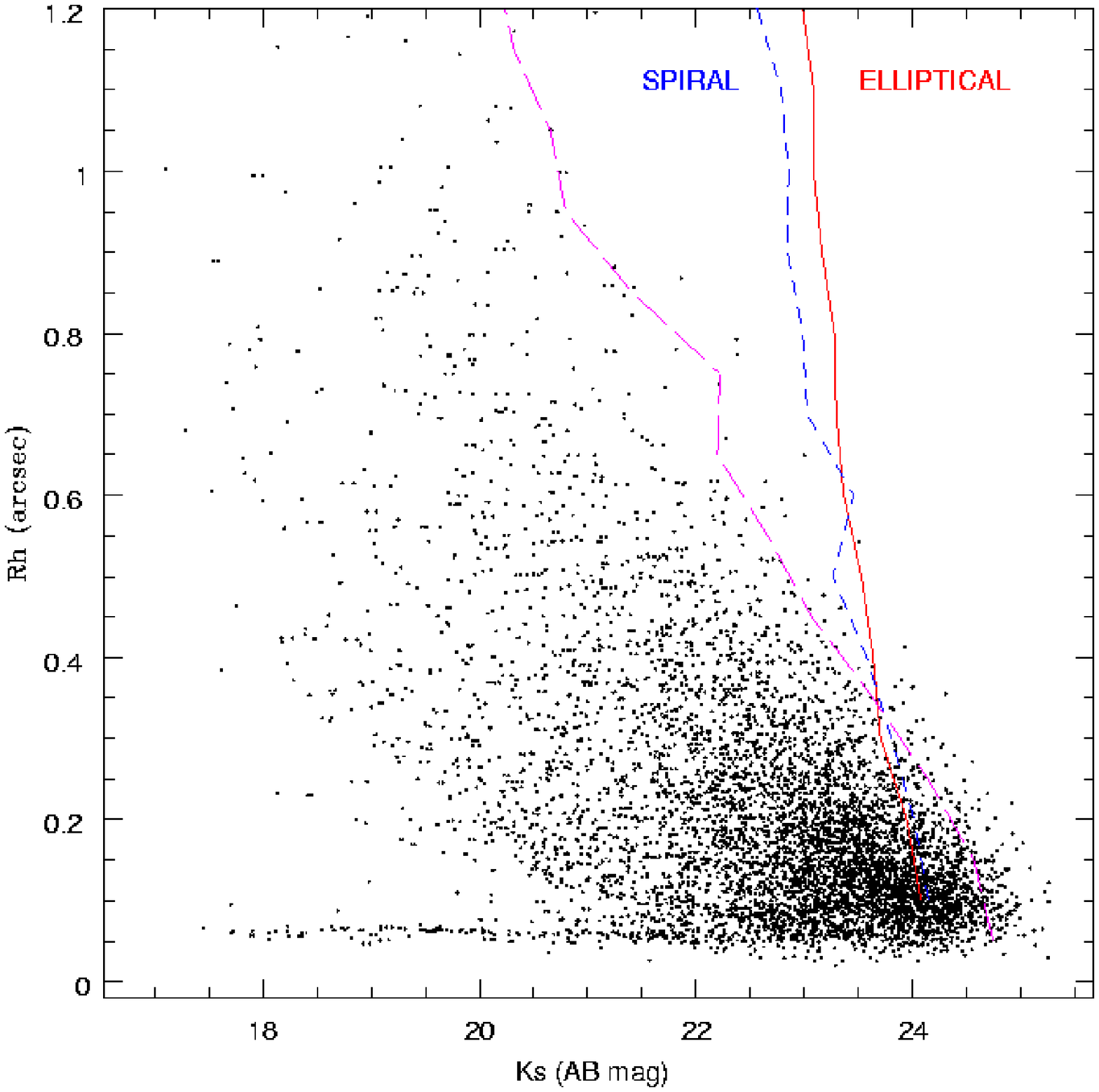}
\caption{
The half-light radius of galaxies in the GOODS-CDFS region as a function
of the observed $z$ and Ks magnitudes. The solid line indicates the
completeness at $90\%$ value for elliptical galaxies, while the short-dashed
line shows the same for spiral galaxies.
We simulated 100 galaxies for each bin of magnitude, half-light radius, and
bulge/disk ratio: the magnitude range is from 20 to 30 in steps of 0.5,
the half-light radius range from 0.1 to 1.5 in steps of 0.1 arcsec, and the
B/D from 0.3 to 0.9 in steps of 0.1 for ellipticals or from 0.02 to 0.08
in steps of 0.01 for spiral galaxies. The completeness as a function of
magnitudes and
half-light radii is the fraction of galaxies recovered by our detection
procedure, averaged over the whole bulge/disk ratio interval.
Long-dashed lines represent the fainter envelope of the observed
galaxies. The completeness limit is given by the intersection of the
latter curve with the brightest segments between simulated ellipticals
and spirals.
The plume of observed objects with bright magnitudes and with small
half-light radius are stars.}
\label{maglimZfig}
\end{figure}

Figure \ref{maglimZfig} shows the 90\% completeness level resulting
from such simulations in the $z$ band, for both elliptical and spiral
galaxies of different half--light radii and bulge/disk ratios.
Comparing these selection
functions with the observed distribution of magnitude and size for
real objects in the GOODS field, we derived the exact value of the
total magnitude $z$ at which our catalog is complete.
Table \ref{maglimtab} provides the total magnitudes to build a complete
catalog of sources in the GOODS area.

The same exercise can be done in the Ks band images, where
there are 6 main areas, with a broader range of magnitude
limits. A typical value of 0.06 was used as the r.m.s. value for the
simulations in the Ks band, corresponding to a limiting magnitude of
27.0 in a 1 sq. arcsec area at 1 sigma. This case corresponds to an
exposure time of 7 hours, and defines the
complete zone at a magnitude of $Ks=23.8$.

\section{Colour estimate}

As described in Sect. 2, the GOODS data span a wide range of image
qualities, both in terms of resolution/sampling and in terms of
limiting depth.
In this case, it is difficult to design the optimal strategy.
On the one hand, one would like to make full use of the spatial and
morphological information contained in the highest quality images,
limiting the loss of information occurring when
lower resolution images are included as much as possible.
This requirement would call for
restricting the area over which photometry is done. On the other hand,
an unbiased estimate of the colours is essential for most of the
scientific application of the GOODS dataset (e.g. photometric
redshifts or analysis of the multiwavelength spectral
distribution). Since most of the detected objects are galaxies that
may exhibit colour gradients due to a change in the morphological
properties across the wavelengths, one would like to ensure that a
properly wide area is taken for each object. Finally, an optimal S/N
is required to improve the accuracy of photometric redshifts.

To match these discordant requirements as closely as possible, we
adopted two different techniques to accurately measure colours, which we
adopt in the ACS and in the ground--based images, respectively. These
methods are described in this section.

\subsection{Colours in the ACS images}

Colours in the ACS images were computed by using the {\it
isophotal} magnitudes as obtained by SExtractor in ``dual
image'' mode.  This procedure ensures that the photometry is computed
at the different wavelengths in the same physical region, defined by
the isophotal area in the ``detection'' image (here, the $z$ band).
In the high resolution ACS images, the isophotal area follows the
apparent size of the objects and is relatively insensitive to the
effects of nearby contaminants. For these reasons, it has also been
used by a number of works dealing with similar data set (\cite{labbe02,
vanzella01,cimatti02}).

However, we also explored alternative solutions, all based on the
``dual image'' mode of SExtractor. A fixed aperture that is commonly
used in the ground--based photometry of faint objects is inapplicable for
the large dynamical range of the GOODS images: the typical isophotal
radii range from 2 to 20 pixels, but can be as large as 150 pixels.

``Kron'' magnitudes (``MAG\_AUTO'' in SExtractor slang) are
typically estimated on areas that are larger than isophotal ones, and
might be less sensitive to colour gradients. We verified on bright,
as well as on faint, isolated objects that they provide colours that are
(on average) identical to isophotal ones. However, they may be
contaminated by neighbouring objects and typically have a lower S/N
than isophotal, since they include more pixels of low S/N.

We also attempted to use an ``optimal'' circular aperture,
defined as the aperture which maximises the S/N of each object in the
detection image. Although promising, we verified that it is difficult
to provide a robust estimate of such an aperture, since a nearby galaxy may
easily contaminate the automatic estimate of the S/N of faint
galaxies.
Eventually, we compared the quality of the photometric redshifts
obtained with different colour estimators, applied to objects with a
spectroscopic redshift, finding that isophotal colours provide a
slightly better result than other choices.
On the basis of these results, we adopted isophotal colours for
all the objects in the sample, on all the ACS images.  For
ground--based images, we developed a more complex technique that
we discuss below.

\subsection{PSF-matched colours in ground--based images}

To measure colours between ACS and ground--based images, we
specifically developed and adopted a ``PSF--matching'' code designed
to work especially for faint galaxies, which makes it possible to accurately
measure colours in relatively crowded fields, while making full use of
the spatial and morphological information contained in the highest
quality images.

The technique that we adopted has been introduced for
the first time by the Stony-Brook group to optimise the analysis of
the $J$, $H$, and $K$ images of the Hubble Deep Field North. The method
is described in \cite{fernandez99}, and the catalog
obtained has been used in several scientific analyses of the HDFN,
both by the Stony-Brook group (e.g. \cite{lanzetta99,phillipps00}),
as well as by
several other groups. More recently, a conceptually similar method has
been developed by \cite{papovich} to deal with similar
problems existing in the first release of GOODS data set.

\subsubsection{The ConvPhot algorithm}

A full description of our code that we name ConvPhot is beyond the
aim of the present work. Since we make it publicly available, we
refer the reader to its accompanying manual (De Santis et al.,
in preparation) for a full
description. We recall the basic principle here and highlight the
possible systematic effects, the strategy that we adopted to minimise
them, and the validation tests we performed.

\begin{figure}[t]
\begin{center}
\includegraphics[height=.4\textheight]{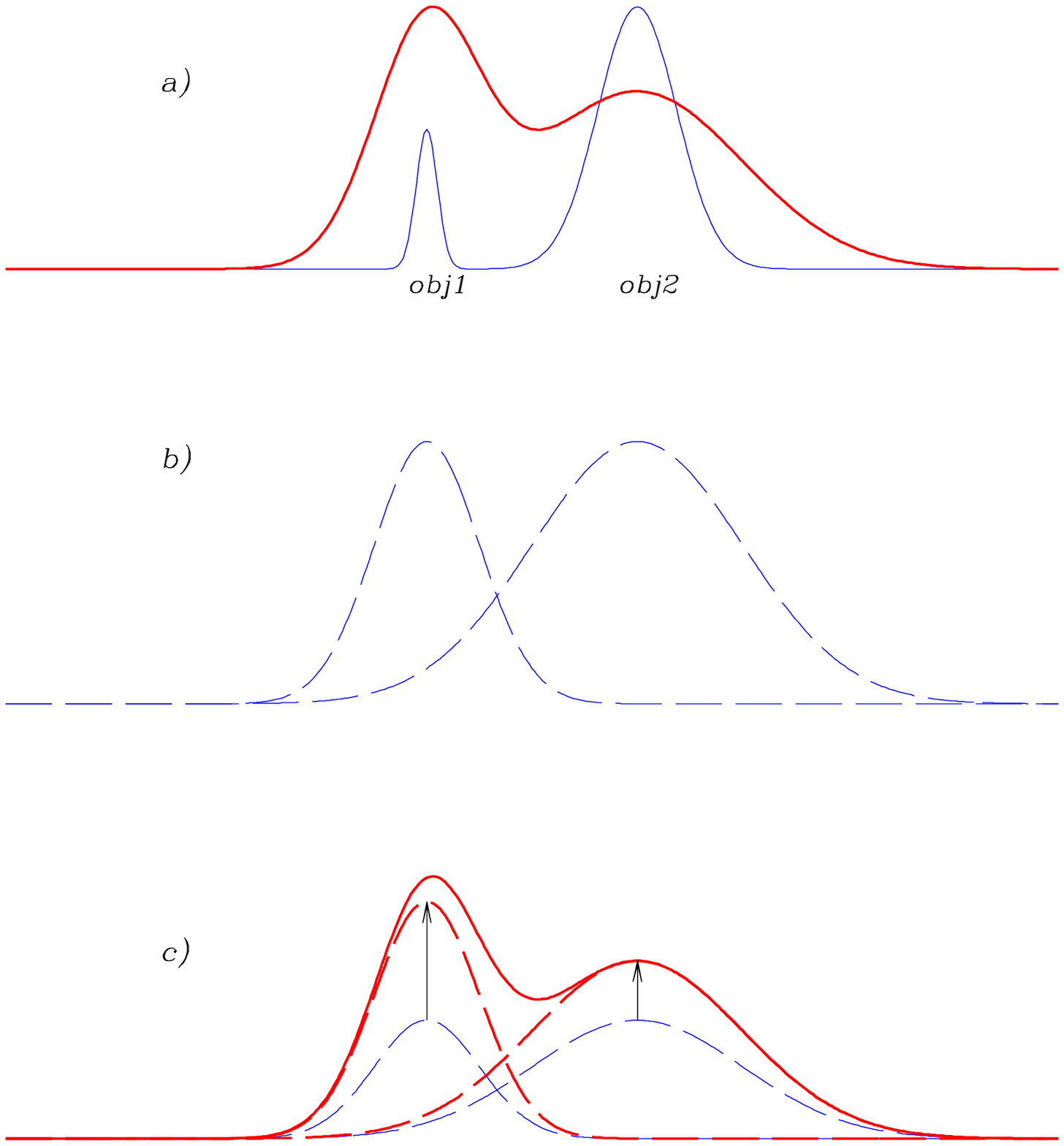}
\end{center}
\caption[]{A schematic representation of the ConvPhot algorithm.

{\it a)}. Two objects are clearly detected and separated in the high
resolution \di~ image (blue, solid-thin line). The same two objects are
blended in the low resolution ``\mi'' image (red, solid-thick line)
and have quite different colours.  {\it b)}. The two objects are
isolated in the high resolution \dii~ and are individually smoothed to
the PSF of the \mii~ to obtain the ``model'' images.  {\it c)}. The
intensity of each object is scaled to match the global profile of the
\mii. The scaling factors are found with a global $\chi^2$
minimisation over the whole object pixels.}
\label{conv}
\end{figure}

Conceptually, the method is quite straitforward and can be better
understood by looking at Fig. \ref{conv}, where we plot the case of
two objects that are clearly detected in the {\it``detection image''}
but severely blended in the {\it``measure''} one. The procedure
followed by ConvPhot consists of the following steps:\\ {\it a)} Each
object is extracted from the {\it``detection image''}, making use of
the parameters and of the isophotal area defined by SExtractor.  Since
the latter typically underestimates the actual object size,
the SExtractor isophotal area is expanded by an amount that is proportional to
the object size.  \\ {\it b)} Each object is individually filtered to
match the {\it``measure''} PSF and normalised to unit total flux: we
refer to the resulting thumbnails as the ``model profiles'' of the
objects.\\ {\it c)} The intensity of each ``model'' object is then
scaled in order to match the intensity of the object in the
{\it``measure''} image. The free parameter for this scaling ($F_i$)
is computed with an $\chi^2$ minimisation over all the
pixels of the images, and all objects are fitted simultaneously to
take the effects of blending into account between nearby objects.
Although the number of free parameters that is equal to the number of
identified objects is quite large, the resulting linear system is
very sparse and can be efficiently solved using standard numerical
techniques.

\subsubsection{Systematics in the PSF--matching}
As can be appreciated from the example plotted in Fig. \ref{conv}, the
main advantage of the method is that it relies on the accurate
spatial and morphological information contained in the ``detection'' image
to measure colours in relatively crowded fields, even in the case where 
the colours of blended objects are markedly different. 

Still, the method relies on a few assumptions that must be well
understood and taken into account. In particular, morphology and
positions of the objects should not change significantly between the
two bandwidths. Also, the depth and central bandpass of the {\it
``detection''} image must ensure that most of the objects detected in
the {\it``measure''} image are contained in the catalog. The
objects that are deeply blended in the measure image should be
separated well on the detection image.

In practice, it is unlikely that all these conditions are satisfied in
real cases. In the case of the match between ACS and ground-based $Ks$
images, for instance, very red objects may be detected in the Ks band
with no counterpart in the optical images, and some morphological
change is expected due to the increasing contribution of the bulge in
the near--IR bands. Also, in the case where the pixel size of the
{\it``measure''} image (i.e. ISAAC or VIMOS) is much larger than
in the {\it``detection''} one (ACS $z$ in our case), the actual limit
intrinsic accuracy in image aligning may lead to systematic errors.

To characterise and deal with these possible uncertainties, we
included several options and fine--tuning parameters in the code to
minimise the systematics involved, and used extensive simulations to
choose them in an optimal way. We briefly describe them here, referring
to a separate paper for a full description.

\smallskip

{\em Missed flux in the outer area}\\ The Segmentation image provided by
SExtractor is usually smaller than the actual object size, since it is
limited to the threshold used to detect that source. As a result, the
object profile and total flux are incorrectly recovered, which may
result in a systematic bias in the fitted colours.

To minimise this effect, we expanded the segmentation produced by
SExtractor by an amount that is proportional to the object size, with
a minimum area.  A minimum value for the area of extended segmentation
is useful in the ACS images of the GOODS-CDFS, where extended but low-surface
brightness objects are detected only due to the bright nucleus,
and the isophotal area is limited only to the brighter knots.  For the
GOODS-CDFS $z$-band image we dilate the resulting segmentation such
that the objects are doubled in linear dimensions, preserving their
shape, and in the case of very small objects, the area is set to an
equivalent 1 arcsec (diameter) circular aperture.

{\em Blended sources in the detection image}\\
The {\em ConvPhot} algorithm is ideally suited to photometry of
blended objects on the measure image, but requires that the same
objects in the detection image should be well defined and their
profiles not be distorted by noise. In the practical case, even in the
GOODS-ACS images, there are galaxies blended in the ACS images or
faint objects, brighter than the detection limit but still affected by
noise in their shape/profile.  To overcome this problem, we introduced
an option in {\em ConvPhot} to carry out the fitting procedure only
on the central part of the profile: the fit is restricted only to the
pixels where the detection image is above a given relative threshold.
This ensures that the fit is carried out only where the signal-to-noise
of the model (detection image) is high and avoids
contamination from nearby sources, which cannot be perfectly modelled
in the detection image.  A threshold of 0.5 is used for the GOODS dataset,
after extensive simulations.

\smallskip

{\em Alignment errors}\\ The fitting procedure is extremely sensitive
to alignment errors.  In the simple case the object having a 2-D
Gaussian shape, it is easy to show that the resulting flux is 
systematically underestimated  by a factor $f = exp(-\frac{3}{4}
\frac{\Delta r^2}{\sigma^2})$, where $\Delta r$ is the offset in pixel
of the central position. When ground--based images are combined to HST
images with excellent sampling, this effect is non-negligible. In
the case of the GOODS ACS data, for instance, the ACS pixel size is
0.03'', which is often smaller than the residuals of the alignment of
IR ground--based images. For an alignment error of 1 pixel (in the
\dii), a figure that can be quite typical or even optimal when
combining ground--based and HST images, the resulting underestimate
may be about 3\%.

As a first way out, we included an option in \cp~ to re-centre any
object before minimisation. In this case, the centre of each object
is internally computed in the model, as well as in the measure image,
and the measure is re-centered to the model image before minimisation.
Since the centre determination may be noisy for faint objects, the
user can set a limit on the S/N of the objects (in each image) for
this operation. We use the re-centering options only for WFI U band images,
only for sources with $S/N\ge 15$ both in model and measure images.

\smallskip

{\em Variable FWHM or object profile}\\ Another source of uncertainty
may result from a variation in the object profile from the \di~ to the
\mii, i.e. when the profile in the model image is markedly different
from the real profile in the measure image,
after the smoothing with the PSF transformation kernel.
This can be due to either a physical change of the object
profile (as due, for instance, to a more prominent bulge in the IR) or
to an incorrect estimate of the PSF transformation kernel. In this
case, the resulting flux can therefore be either under- or
over-estimated, depending on the sign of the error in the PSF
estimate. An error of 10\% in the object FWHM will result in a 5\%
error in the output flux.

The small systematic effects that have been described above, or others
resulting from different sources, can be efficiently corrected by
taking the flux in residual image into account. For this purpose, we
included an option in the code to compute the total residual flux
contained in the segmentation area of the original frame. 
This ensures the systematics of colour estimations to lower down, as shown in
the next paragraph and in the \cp~ paper (De Santis et al.,
in preparation).

\subsubsection{Validation tests}

We performed several validation tests on the ConvPhot code,
during the debugging phase to estimate the efficiency in the
correction for systematics.  The most obvious involved the use of
simulated images, with a range of luminosities, PSF, and morphologies,
by which we verified that the code is computationally correct
(De Santis et al., in preparation).

\begin{figure}
\includegraphics[width=9cm]{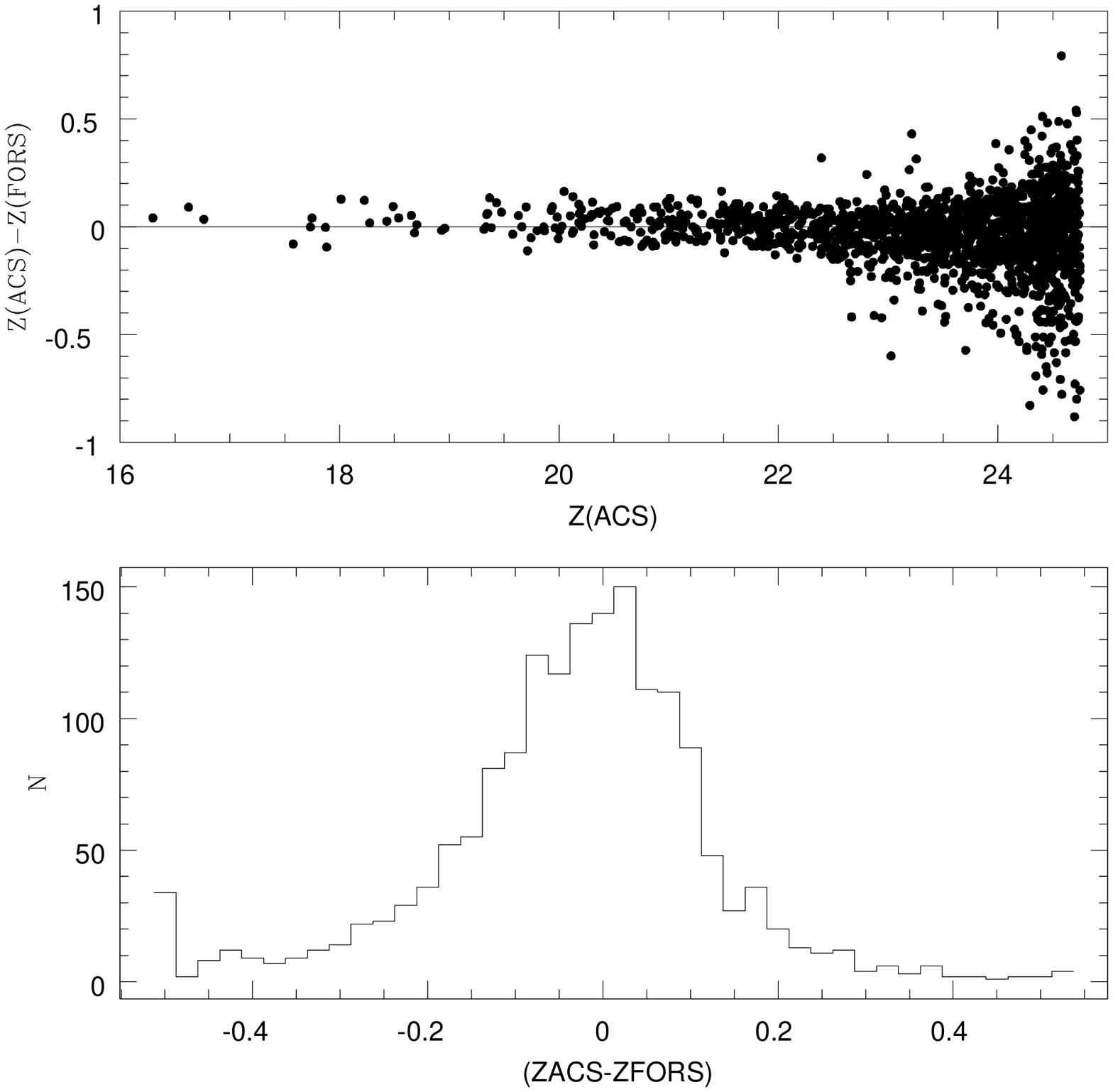}
\caption{The $z_{ACS}-z_{FORS}$ colour as a function of the magnitude of the
objects detected on the K20 survey (upper panel). The histogram of
this colour (lower panel) shows that the \cp~ magnitude determination
is not biased and is compatible with aperture photometry.  }
\label{FORS}
\end{figure}

Simulations, however, may not fully reproduce the complexity of real objects
and data. To obtain a more stringent and independent test, we made
use of the $z$ band FORS image of the CDFS obtained within the K20
survey (\cite{cimatti02}).  Here, we used \cp~ to obtain a new
estimate of the $z_{ACS}-z_{FORS}$ colour in the FORS image of the K20.
Since the two $z$ filters are fairly similar, all objects should have
a null colour, barring variable objects. In practice a small offset (0.035 mag)
between the two bands persists still even at bright magnitudes, probably due
to a different response for the ACS and FORS instruments. We limit the
comparison to $z\le 24.75$, the limit of FORS $z$ band at  $S/N=10$.
Due to the brighter magnitude limit of the FORS image, the error in
the $z_{ACS}-z_{FORS}$
colour is dominated by the error on the FORS z-band magnitude estimate.
The results of this comparison are
shown in Fig. \ref{FORS}, and show that the
\cp~ software is not biased in the colour determination.

\subsection{The final catalogs}

The direct output of a \cp~ run is the scaling parameter $F_i$ for
each object. Since the model profile $P_i$ for each object is normalised to
unit flux, the resulting total magnitude in the \mii~ is simply $-2.5
\log(F_i) + ZP_m$, where $ZP_m$ is the zeropoint of the \mii~
itself. Based on our tests, we concluded that this total
magnitude is a reliable measure of the actual total flux of the
objects, somewhat less prone to systematic effects than the Kron
magnitudes computed by SExtractor. However, these total magnitudes can
hardly be compared with the SExtractor magnitudes of the \dii, so that
reliable colours cannot be obtained directly. To this end, we used
the total flux $D_i$ measured by \cp~ itself in the \dii~ and used it to
normalise the object profile $P_i$.  The resulting flux ratio is
therefore $Flux(measure)/Flux(Detection) = F_i / D_i \times
10^{-0.4(ZP_m - ZP_d)}$, where $ZP_d$ is the zeropoint of the
\dii. This flux ratio, or the equivalent magnitude colour,
$m_{measure}-m_{detection} = -2.5 \log(F_i / D_i) +ZP_m - ZP_d$ is the
final colour that we used in the following. All colours were finally
normalised to the total $z$ magnitude to obtain self--consistent
magnitudes at all wavelengths.

As stated above, the final colours were obtained with \cp.  For
the $J$, $H$, and $Ks$ images, we resampled and aligned each
individual VLT--ISAAC image to the ACS $z$ mosaic that we used for
object detection, and used the $z$ image, smoothed with the appropriate
convolution kernel, as the ``model'' image. In the
case of the $U_{35}$, $U_{38}$, and $U_{VIMOS}$ images, we followed the
same procedure, after cutting the original $U$ images into smaller
pieces (since \cp~ can operate only with small size images, due to memory
limitations), but using the ACS $B$ image, smoothed with the appropriate
convolution kernel,
as a ``model'' to minimise the effects of wavelength--dependent
morphologies.  Because of the geometry of IR images, several objects
were detected in more than one IR pointing. In these cases, an
optimally weighted average was used to estimate the final
colour. In all cases, we conservatively dilated the
``Segmentation'' image by a factor of 4, with a minimum area after the
dilate of 800 pixels (corresponding to a PSF with 1 arcsec of diameter),
and applied a fitting threshold of 0.5 to
minimise the effects of nearby contaminants. For ISAAC and VIMOS
images, we did not use the centering option, while for the $U_{35}$ and
$U_{38}$ photometry we applied a centering with \cp~ for objects with
$S/N\ge 15$.

The same technique was applied to the Spitzer images. We are
aware that the properties of the PSF of the IRAC instrument is not
fully characterised (FWHM variation across the FoV), and the much lower
image quality (the FWHM ranges
from 1.6 to 2.0 arcsec, see Table \ref{summaryGOODS}) makes the whole
analysis more difficult and uncertain. For this reason, we regard the
colour estimate in the
Spitzer bands as more uncertain than in the other wavelengths, and we plan to
address it in more detail in a future paper where we will also deal with the
Spitzer-selected samples.

For the Spitzer bands, we used {\em ConvPhot} to carry out the
photometric analysis of the GOODS sources detected in the $z$
band.  We dilated the ``Segmentation'' image by a factor of 4,
with a minimum area after the dilate of 800 pixels, and applied a
fitting threshold of 0.5 to minimise the effects of nearby
contaminants. For IRAC images, we haven't used the centering option.

\begin{figure}
\includegraphics[width=9cm]{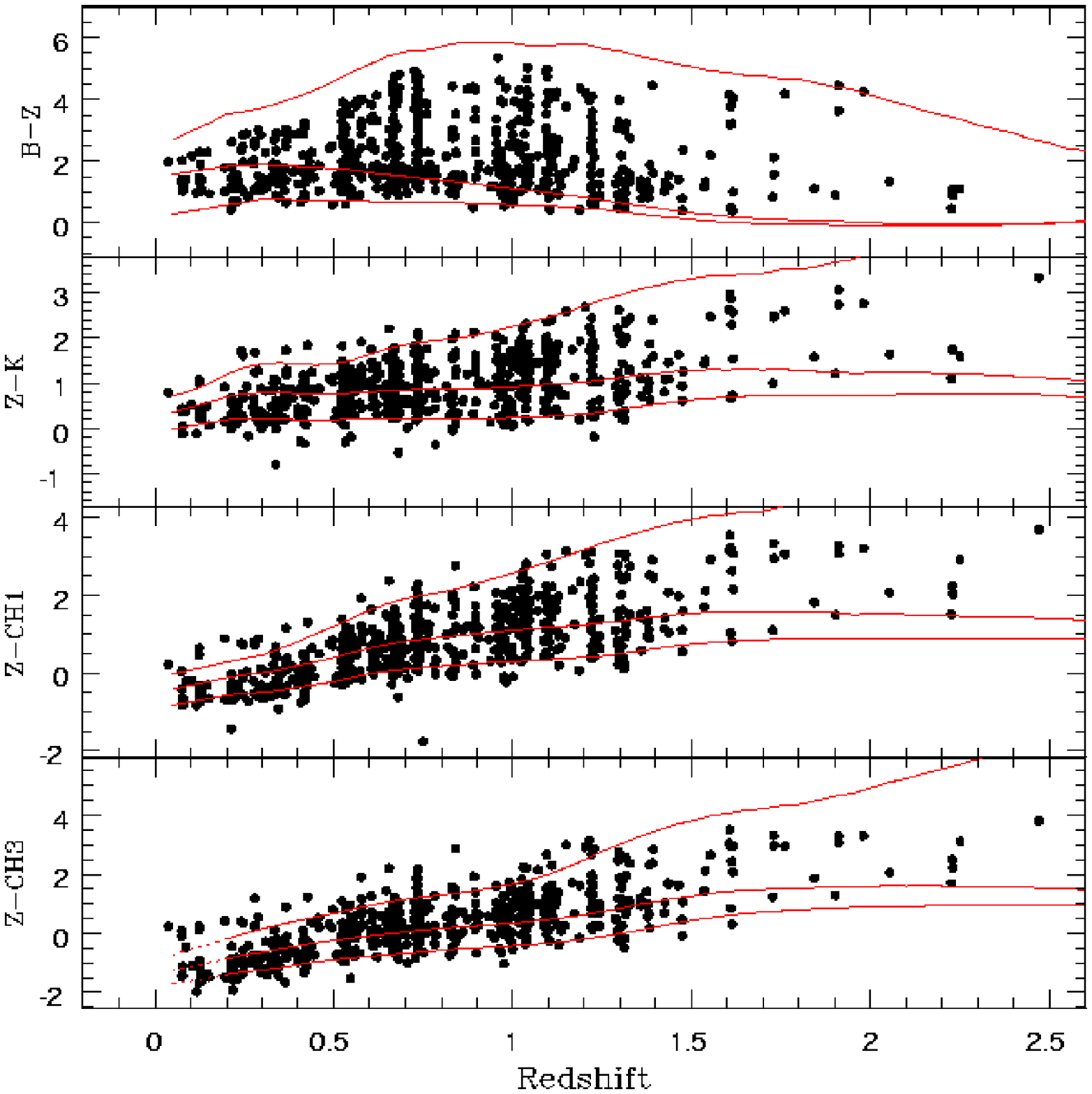}
\caption{Observed colours as a function of redshift for the
objects with spectroscopic redshifts in the GOODS-S field.  The solid
lines are (from top to bottom) the colours predicted by the \cite{bc03}
code for the following cases: (i) a maximally old model,
obtained assuming an exponential history of star--formation with
timescale 0.1 Gyrs started at $z=10$ with metallicity $Z=2.5 Z_\odot$; (ii)
a similar model but with an exponential timescale of 5 Gyrs and 
an actively star-forming and non--evolving model with a constant
star--formation rate and age of 1 Gyr.
Dotted lines represent the redshift range in which the
colours are not reliable due to the incomplete modelling of dust emission
at $\lambda \ge 5.5\mu m$ rest frame (see Sect. 7).
}
\label{zcolour}
\end{figure}

Figure \ref{zcolour} shows the colours obtained for all galaxies with
spectroscopic redshift in our sample (the sources of the spectroscopic
data adopted are listed in Sect. 6) in a few selected bands.
The corresponding range obtained from the \cite{bc03} program is also shown
for two extreme models, a very young, star--forming
model, and a maximally old model with a short e-folding star-formation
timescale. Most of the objects lie within these boundaries at the
different redshifts, providing a reassuring quality check on the
overall photometry. We verified that such an agreement still
holds in the other bands.

\section{The Ks-selected catalog}

The catalog produced using the $z$ band of ACS as the detection image is not
complete in the Ks band magnitude. In principle, this could be
obtained by using the $Ks$ images as detection images, and computing
colours in the other bands. The catalog obtained, however, would not
benefit from the higher spatial resolution of the ACS images, and the two
catalogs would not be homogeneous. To prevent this, we followed a
different approach of identifying the objects missed in the
$z$--selected catalog in the so-called ``Drop'' images produced by
ConvPhot in the Ks band, i.e. the residuals of the fitting procedure.

\begin{figure}
\includegraphics[width=9cm]{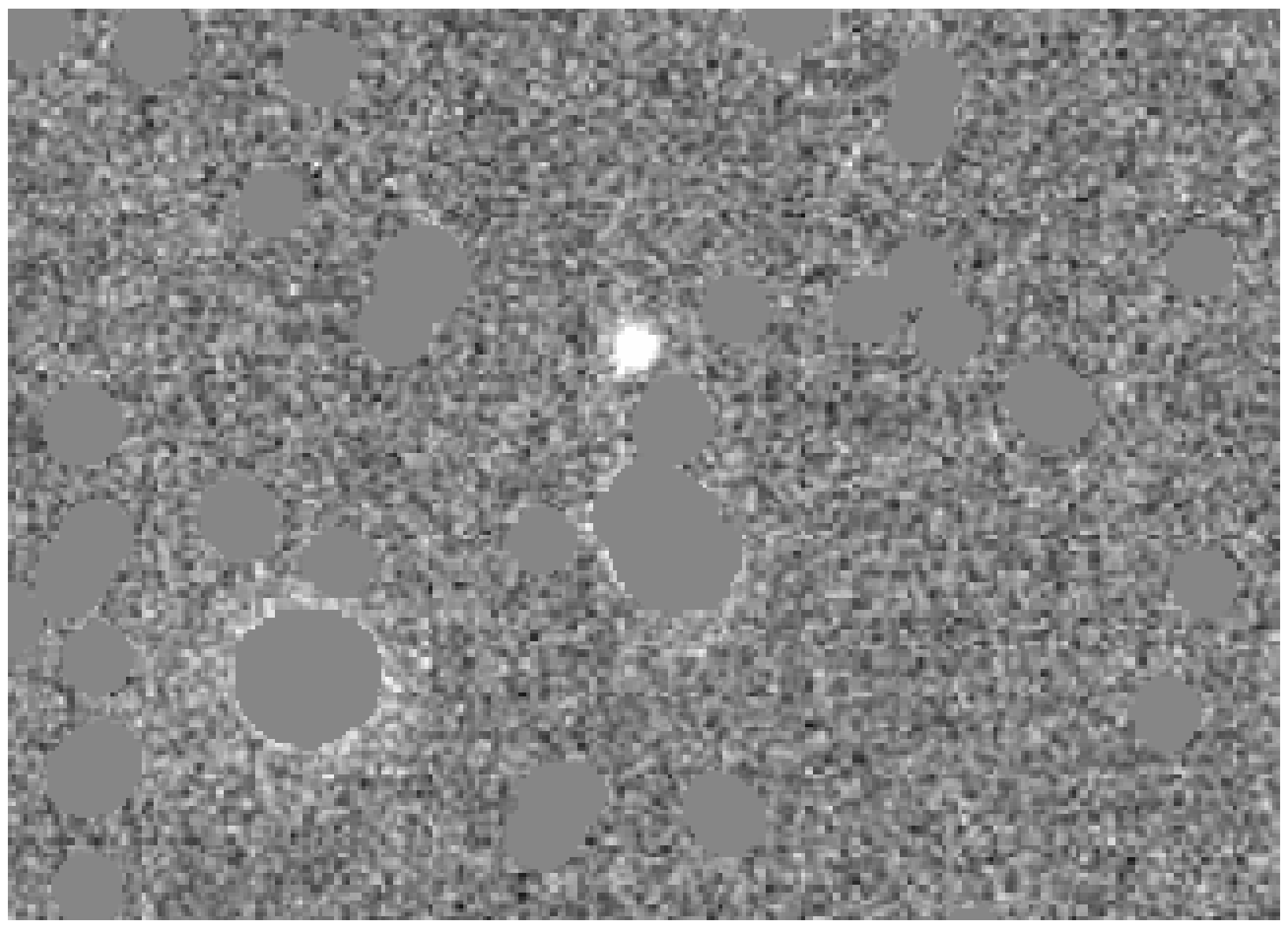}
\caption{A so-called ``Drop'' image produced by ConvPhot in the Ks band.
The name derives from the fact that each $z$-band detected object that is
fitted by ConvPhot is multiplied by zero (drop) in order to leave only
those Ks bright galaxies not detected in the $z$ band.
}
\label{zdrop}
\end{figure}

An example is shown in Fig. \ref{zdrop}, which shows that there are
objects that are not detected in the $z$ band but bright in the Ks band.
To identify all these missed objects, we first carried out a run
with ConvPhot with a dilate parameter equal to 0 and using small
smoothing kernels in order to fit only the central part of the $z$ 
detected objects and not to lose Ks bright galaxies close to  $z$
detected objects.  In this way the Drop images have a minimal area
covered by ``drops'', and the completeness of the Ks selected objects
is enhanced.

We ran SExtractor on the Ks ``Drop'' images, producing a complete
catalog, whose depth will follow the exposure maps described
above.  On the basis of the simulations described above, it is
typically complete down to $Ks=23.8$ magnitude (AB) at the 90\%
completeness level.  As total Ks band magnitude, we used the BEST
magnitude of SExtractor.

The extraction of the colour information for these Ks selected objects
is not trivial. We calculated the colours in an isophotal area defined
by the dilated segmentation of the Ks band objects. In order to
provide an unbiased estimate for the colours, we resampled the ACS
images to the Ks resolution using the convolution kernel adopted by
\cp.  The total magnitude in these bands is computed using the Ks as
reference and the colour term as an additive effect:
$B_{tot}=Ks_{tot}+(B-Ks)_{iso}$, where $(B-Ks)_{iso}$ is the colour
computed in the ``extended'' isophotal area.  For the other bands (U
and IRAC), where the FWHM is larger than Ks, we resampled the Ks band
image to match the measure image PSF.

At the end we detected 196 objects not present in the $z$
selected catalog.  The nature of these objects is typically twofold:
they can be low-surface brightness galaxies in the ACS $z$ band
images, extended objects that escape from detection criterion in $z$,
but with bright total magnitudes detected in the Ks band because
of a lucky combination of pixel-size and seeing effects (151,
77\%). The other type is constituted by objects that are not detected
in the ACS image because they are really $z$-dropout objects,
basically EROs at $z\ge 1-2$ (42, 21\%). Only three objects are not
part of these two categories: bright objects in the $z$ and
Ks bands that are not detected in $z$ because they are on a spike of a
bright star (2 objects) or in the halo of a bright and extended galaxy
(1). We recover them from the Ks Drop images.

We merged the $z$-selected catalog with the Drop Ks selected
objects in order to build a total catalog of objects that can be
considered complete down to $z\simeq26.0$ and $Ks\simeq24.0$
simultaneously.  In the catalog, $ID\le 20000$ indicates $z$-detected
objects, while $ID\ge 30000$ points to ``drop'' Ks selected
objects. We also computed, as described in \S
2.4, the limiting magnitudes (at 1 $\sigma$ and in 1 sq.
arcsec. area) in $z$ and Ks bands for the Ks detected objects, in
order to have the same quantities of the $z$-detected catalog to
ensure a given homogeneity to the total catalog. At the moment there
is no spectroscopic information for the Ks drop detected galaxies,
since they are not bright enough (in surface brightness limited
samples) in the optical bands or not present in previous ACS-based
catalogs used for spectroscopic identifications.

\section{The spectroscopic catalog}

The GOODS-CDFS field has been the target of a number of spectroscopic
surveys that paved this area with unprecedented accuracy.
The various catalogs surveyed this field for different reasons:
the COMBO-17 survey (\cite{combo}) needed spectra to calibrate
their photometric redshifts over a wide and relatively shallow area;
the CXO survey (\cite{cxo}) is a spectroscopic follow-up of X-ray
sources in the Megaseconds CDFS; and the K20 survey
(\cite{k20}) provides the spectra for 327 sources with $K\le
20(Vega)$. After the public release of the ACS GOODS images, two
spectroscopic surveys were released: the GOODS V1.0 spectroscopy
(\cite{vanzella05}) and the VVDS survey (\cite{vvds}) with the aim of
providing a comprehensive census of bright objects in the GOODS
field. Recently the ESO-GOODS team released a
Master\footnote{http://www.eso.org/science/goods/spectroscopy/CDFS\_Mastercat/}
catalog of spectra in the GOODS-CDFS area, which summaries
all the redshift information of the previously cited catalogs,
with additional redshifts from sparse surveys.

The information provided by these surveys is basically coordinates
and redshifts. The classification (Star, galaxy, or AGN) and a quality
flag for the spectrum are provided by all of the catalogs, but are not
present in CXO and Master catalogs. The quality flags of the various
catalogs are not homogeneous, so we tried to define a
homogeneous classification for all the catalogs. For the
K20 catalog, the quality of the redshifts is $q_{K20}=1$ for the secure
redshifts and $q_{K20}=0$ for the uncertain ones. For the GOODS spectroscopy,
the quality flag goes from A to C, towards a decreasing S/N of the
spectra. The VVDS uses a reverse scheme when compared to our way of
quantifying the quality of the spectra, with $q_{VVDS}=4$ for the highest
quality to $q_{VVDS}=1$ for the lowest one. We define four
classes of quality flags, from 0 to 3, where qz=0 indicates the best spectra,
with secure identification and qz=3 the most noisy spectra.
For the other spectroscopic catalogs, where the quality flag is not
available, a qz=1 is defined a priori.
Only for the K20 and GOODS spectroscopic
surveys is there the distinction between early or emission line galaxy
(or a combined spectrum). For the other catalogs, the classification is
not homogeneous and is not subdivided in finer classifications such as
early, emission line, or composite spectra. In order to have a
simple classification scheme, we divided all the objects with spectroscopic
information in these classes: STAR, AGN, GALAXY, EARLY, EMISSION, or
COMPOSITE. The classification EARLY, EMISSION, or COMPOSITE only comes
from the K20 and GOODS spectroscopic surveys.

\begin{figure}
\includegraphics[width=9cm]{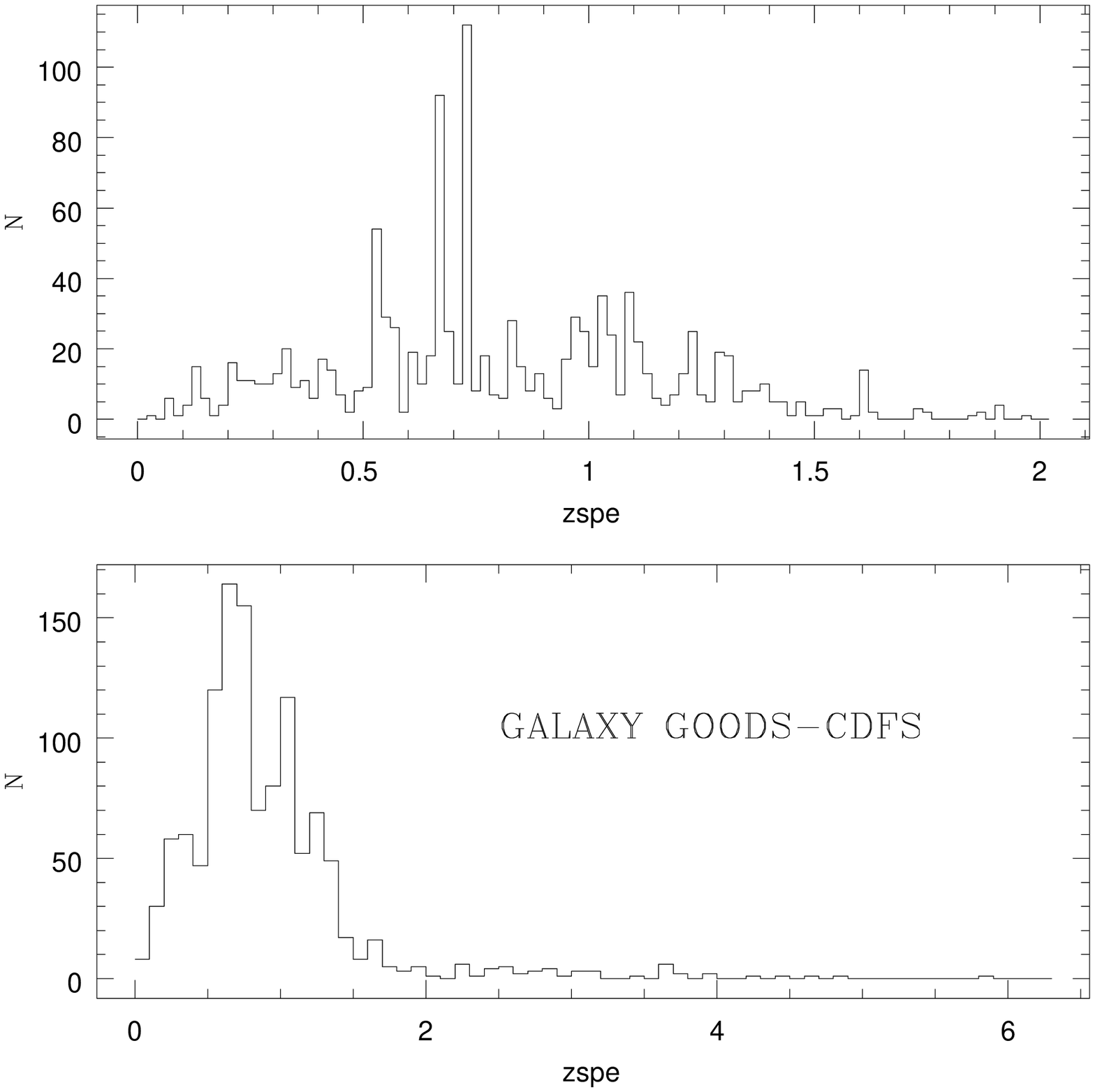}
\caption{The histogram of spectroscopic redshifts for galaxies in the
GOODS-CDFS field.
Three peaks
in redshifts show the presence of large-scale structures like
groups or sheet/wall of galaxies.}
\label{zhist}
\end{figure}

Since in several cases there are multiple identifications for the same
object, we
cross-correlated the catalogs to check for inconsistencies; if there
are two different identifications for the same object, we take the
identification with best quality flag. In the GOODS ACS area there are
currently 1068 known spectroscopic redshifts: Fig. \ref{zhist} shows
the redshift distribution of these objects, in which three distinct
groups/z-peak are visible, corresponding to different groups or sheets
of galaxies.  There are in total 928 galaxies, 72 stars, 68 AGNs, and
QSOs that are already known. In the galaxy sub-sample, only 668 have a
secure redshift, while the remaining are not secure. For 428 galaxies
the finer classification (early-emission line) is available, where
``only'' 94 are classified as early and 67 as composite spectrum.

The identification of the spectroscopic catalog with objects in the
ACS images is not trivial, due to the finest resolution of HST
compared to ground-based images, where close-by objects are merged into a
single blob, or because there is a coordinate mismatch, especially for
the K20 and CXO surveys, based on imaging material precedent to the
ACS one.  To overcome this problem/issue, we carried out a cross
correlation in right ascension and declination between the photometric
data and the spectroscopic catalog with a relatively large matching
radius (1.2 arcsec) and divided the objects into two groups: isolated
objects, with a unique identification and blended or ambiguous
objects, where more than one photometric ID is associated with a
spectroscopic data. In the last case the association is checked by eye
and decided also using the magnitude/colour information.

Overall,
less than 10 percent of the GOODS catalog has any spectroscopic
information (1068 out of 14847). For the remaining objects, having a
high photometric quality and a wide wavelength sampling of the SEDs,
we chose the approach of the photometric (\S\ 7) and neural network
redshifts (Vanzella et al., in preparation).

\subsection{Star-Galaxy separation}

\begin{figure}
\includegraphics[width=9cm]{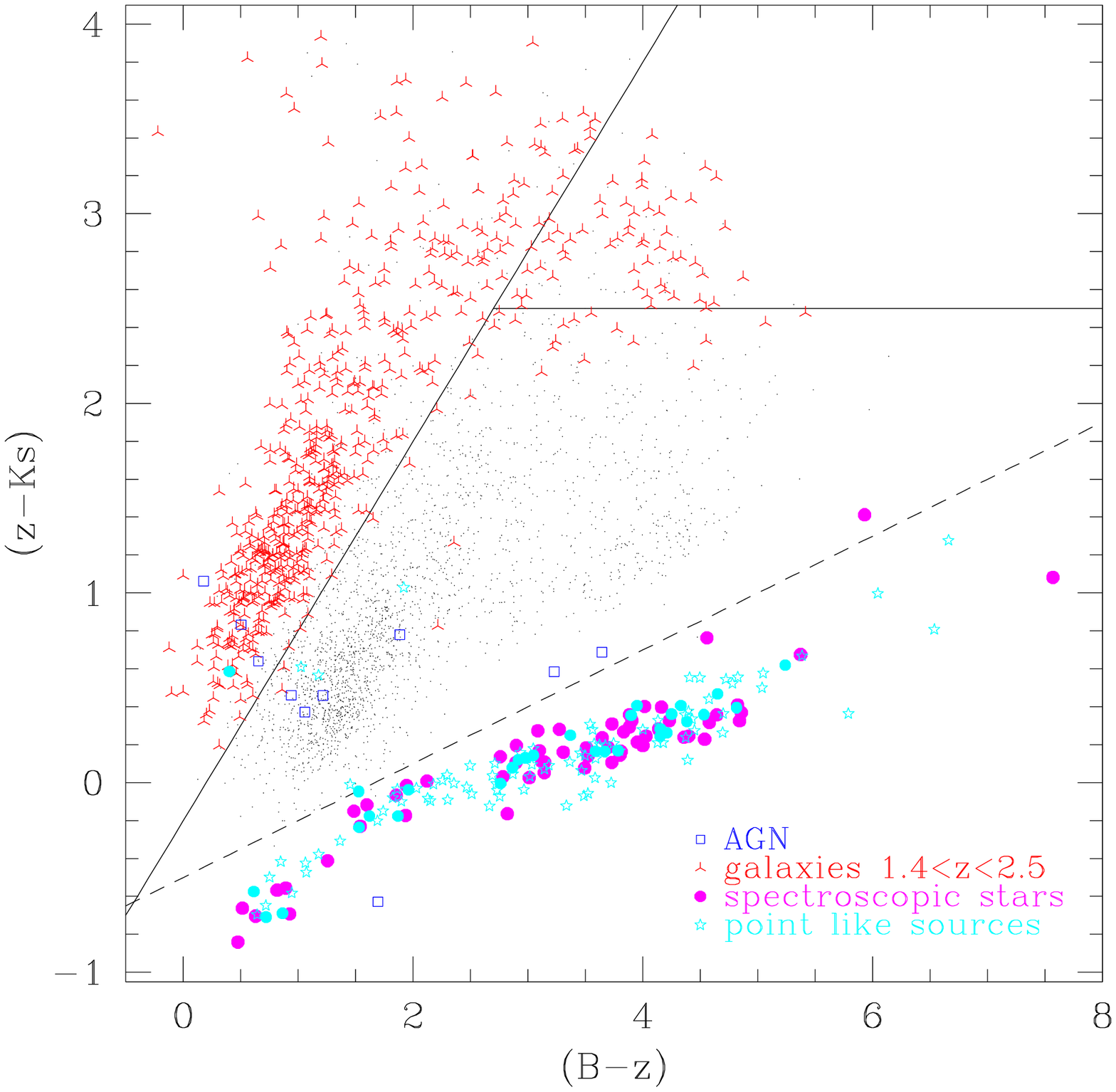}
\caption{
The B-z versus z-Ks colours for objects in the GOODS south. Small dots
are galaxies with redshifts $z\le 1.4$ or $z\ge 2.5$, squares are
spectroscopically known AGNs, arrows indicate galaxies with
$1.4<z<2.5$ (spectroscopic and photometric redshifts), filled points
are spectroscopically known stars, while stars represent objects with
point-like morphology according to SExtractor. Solid lines indicate
the BzK criterion to isolate star-forming (upper left part
of the diagram) and passively evolving galaxies (upper right part of
the diagram) at $1.4<z<2.5$. Objects below the dashed line are stars,
according to \cite{bzk}.
}
\label{bzk}
\end{figure}

We distinguished galaxies from stars and AGNs using
morphological and photometric information, when spectroscopic data
were not available.  First, point-like sources were selected using
the star/galaxy separation flag (s/g) provided by SExtractor in the z
band. We tuned the selection on already known spectroscopic stars:
objects with magnitude $z\le 20$ and $s/g\ge 0.80$, $20<z\le 23$ and
$s/g\ge 0.96$ or $23<z\le 24$ and $s/g\ge 0.98$ were considered
possible AGNs or stars. We used photometric information to check
this criterion, using in particular the ``BzK'' colour criteria of
\cite{bzk}. Figure \ref{bzk} shows the positions of AGNs, galaxies, and
stars in the B-z versus z-Ks two-colour plot. Typically, the point-like
objects selected with the SExtractor parameter were found to lie on
the lower side of the diagram, where stars are expected to be
found. Other objects located in (or nearby) this region, but not
selected morphologically, were visually inspected.  Most of them
turned out to be extended, low-surface brightness galaxies, most
likely at very low redshift (and hence with similar colour to Galactic
stars). The remaining unresolved objects were classified as stars.

\section{Photometric redshifts}

On the multicolour catalog described above we applied our photometric redshift
technique, which was described in detail in \cite{giallongo98}
and \cite{fontana00} (F00 hereafter) which
proved to be extremely successful in the HDFN/S (\cite{fontana03})
and in the K20 data sets (\cite{cimatti02,fontana04}).

In essence, a spectral library of galaxies at arbitrary redshifts is
computed and a $\chi ^2$--minimisation procedure is applied to find
the best--fitting spectral template to the observed colours.  For each
template $t$ at any redshift $z$ we first minimise
\begin{equation}
\chi ^2_{t,z} = \sum_i \left[ {F_{observed,i}-s\cdot F_{template,i}
\over \sigma _i} \right]^2
\end{equation}
with respect to the scaling factor $s$ where $F_{observed,i}$ is the
flux observed in a given filter $i$, $\sigma _i$ is its uncertainty,
$F_{template,i}$ is the flux of the template in the same filter, and
the sum is over the filters used. We then identify the best--fitting
solution with the lowest $\chi ^2_{t,z}$. We refer to the discussion
in F00 for its detailed handling of the non-detection at faint fluxes.
The scaling factor $s$ is therefore applied to the input parameters of
the best--fitting spectrum to compute all the rest--frame quantities,
such as absolute magnitudes or stellar masses.

We tested different choices for the spectral libraries, using
both empirical templates (based on the Coleman et al. data set), as well
as on synthetic models, namely the \cite{bc03} and the
PEGASE 2.0 (\cite{rv97}). We found that the use of
synthetic models results in a better agreement with the spectroscopic
sample available and that, in particular, the choice of the PEGASE 2.0
models minimises both the r.m.s. and the number of outliers
with respect to the BC03 library. In this paper, we briefly describe
the results obtained with the PEGASE 2.0 spectral library to provide
a quantitative description of the quality of our results, and we defer
to a forthcoming paper a more accurate comparison of the performances
of these libraries, which compares these recipes also with the
alternative approach of neural networks (\cite{vanzellaNNz}).

An application of the PEGASE 2.0 code to \fz has been presented by
\cite{leborgne02} on the HDFN dataset. A major advantage of the
PEGASE 2.0 code is that it allows to follow the metallicity evolution
explicitly, also including a self--consistent treatment of
dust extinction and nebular emission.  We adopt the same parameter
grid as in \cite{cimatti02} and \cite{fontana03} which
is very close to the \cite{leborgne02} prescriptions. In detail,
we parameterize the star-formation history by two $e-folding$
star-formation time-scales, one ($\tau_g$) describing the time--scale for
the gas infall on the galaxy and the other ($\tau_*$) the efficiency
of gas to star conversion. By tuning the two time--scales it is
possible to reproduce a wide range of spectral templates from early
types (by using small values of $\tau_g$ and $\tau_*$) to late.
For the earliest spectral type, a stellar wind is also assumed
to block any star-formation activity at an age $t_{wind}$.  Dust
content is followed over the galaxy history as a function of the
on-going star-formation rate, and an appropriate average over
possible orientations is computed. The range of values for $\tau_g$,
$\tau_*$, $t_{wind}$ and the extinction geometry are listed in
Table \ref{tabRV} for a \cite{rana} IMF. For all these models we have
adopted a primordial initial metallicity. Another differences in our
approach compared to
\cite{leborgne02} are that we do not apply any constraint on the galaxy
ages (apart from those set by the Hubble time at each $z$) and that we
have allowed for a finer time sampling in the galaxy ages by halving
the time step with respect to the default value of the PEGASE 2.0
program.

We included the contribution of the nebular emission in the synthetic
library, both in the continuum and in the lines, as allowed
by PEGASE 2.0. At the \lya frequency, however, inclusion of the
corresponding emission line in star--forming galaxies would introduce
a systematic bias, since the observed samples of Lyman break galaxies
at $z\simeq 3$ have a wide distribution of observed \lya equivalent
widths, ranging from absorption to emission systems. For this reason,
we did not apply the \lya emission line as a simple tradeoff, and
we also removed the strong absorption
feature resulting from the stellar features from the synthetic spectrum.

The major difference with the \cite{leborgne02} approach, however, is
due to evidence that the spectral library described above is not
able to reproduce the reddest objects (EROs or similar) that are
typically detected at $z>1$ and that are either passively evolving
galaxies or star--forming dusty objects (Cimatti et al 2002) and that
we detect in our sample. For this reason, we supplemented
the above models with a few templates designed to mimic these
objects.

Passively evolving galaxies were extracted from simple truncated
models, where a starburst with constant star-formation rate is halted
after 0.1, 0.3, 0.6, or 1 Gyr. These models were used only after
the truncation age, so they rapidly become redder than the other models.
These models are able to reproduce redder colours than the
standard models of PEGASE 2.0, particularly in the near and medium
infrared part of the SED, thus fitting the ISAAC and IRAC
observations better.

\begin{table}
\caption[]{Parameters used for the PEGASE 2.0 library. Other models were
added to reproduce the colours of red galaxies at high redshift, as
described in the text.}
         \label{tabRV}
     $$ 
         \begin{array}{p{0.15\linewidth}lll}
            \hline
            \noalign{\smallskip}
            $\tau_g(Myrs)$ & \tau_{*} & t_{wind}  & extinction type \\
            \noalign{\smallskip}
            \hline
            \noalign{\smallskip}
	    100 & 100 & 3000 & spheroid \\
	    100 & 500 & 5000 & spheroid \\
	    500 & 1500 & - & disk, incl-averaged \\
	    1000 & 2500 & - & disk, incl-averaged \\
	    1000 & 5000 & - & disk, incl-averaged \\
	    2000 & 10000 & - & disk, incl-averaged \\
	    2000 & 20000 & - & disk, incl-averaged \\
	    5000 & 20000 & - & disk, incl-averaged \\
            \hline
          \end{array}
    $$ 
\end{table}

Star-forming, dusty objects are instead taken from models with a
constant star--formation rate, following a Calzetti extinction curve
with $0.5\leq E(B-V) \leq 1.1$.
At each redshift, galaxies are allowed any age that is compatible with
the Hubble time at that redshift.

We also added to all models the Lyman series
absorption produced by the intergalactic medium (\cite{madau95}), but
we introduce here a simpler parameterization that uses the DA and DB
values computed from a compilation of data at $z\leq 4$ and
extrapolated at higher $z$ with a simple interpolation to the SDSS
data. In particular we use $DA=[(1.+z)^{2.66838}\times10.^{-2.17281}]$
and $DB=DA^{-0.3576+1}\times 10^{-0.03616}$ at $z\leq 4$, and $DA =
0.187 z -0.247$ and $DB = 1.178 z - 0.133$ at $4.3<z<6.3$, with
$DA=DB=1$ at greater $z$.

\begin{figure}
\includegraphics[width=9cm]{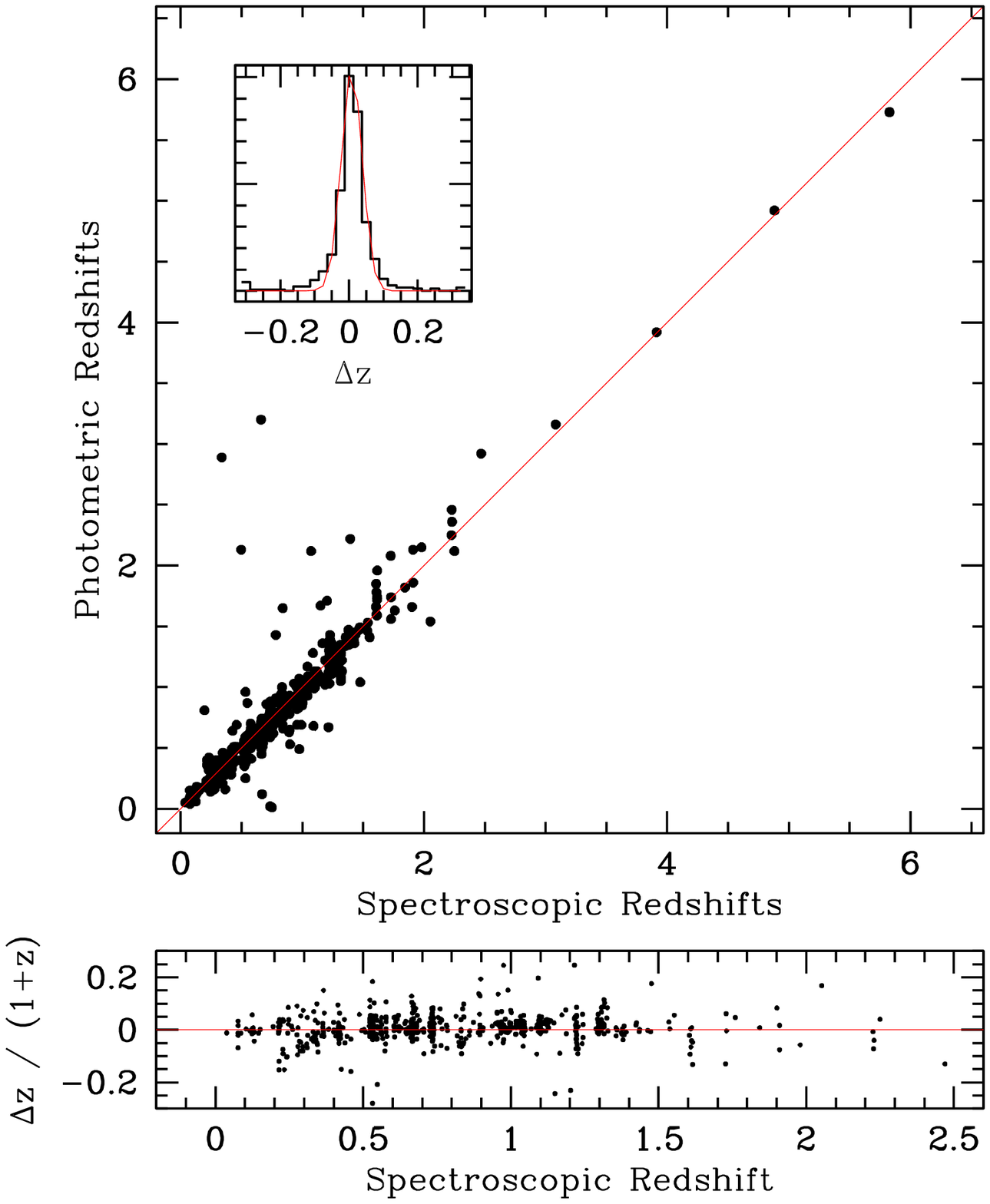}
\caption{
{\it Upper panel}: the relation between the spectroscopic (x-axis) and
the photometric (y-axis) redshift on 668 galaxies with accurate
spectroscopic redshift. In the inset, the distribution of the absolute
scatter $\Delta z = (z_{spec}-z_{phot})$ is shown and compared with a
Gaussian distribution with a standard deviation $\sigma=0.06$ (smooth
red curve). {\it Lower panel}: relative scatter
$(z_{spec}-z_{phot})/(1+z_{spec})$, restricted to the $z<2$ range and
discarding the most discrepant objects in the same sample. }
\label{photoz}
\end{figure}

The most important difference with respect to previous papers arises
from the inclusion of the Spitzer bands, which (at low and intermediate
redshifts) extend the observation to the spectral regions longer than
$3\mu m$ rest frame. The galaxy emission is not dominated by the
integrated stellar population at $\lambda \ge 5.5 \mu m$ (\cite{dale,lu})
but by the different flavours of dust
emission, at least in the case of star-forming galaxies that are not
included in the spectral libraries considered here.  As a simple way
out, we modified our code: any photometric band that is
contaminated by the dust reprocessed emission above $5.5\mu m$ in the
rest-frame is ignored in the fit.

We tested our recipe against the large subsample of galaxies with
secure spectroscopic redshift, which we compiled as described in the
previous section: to this end, we considered 668 galaxies with the
spectroscopic flag equal to 0 or 1.  The results are shown in
Fig.\ref{photoz}. In its upper panel, we plot the $z_{spec}-z_{phot}$
relation to show that we find an excellent agreement between
photometric and spectroscopic redshifts over the fully accessible
redshift range $0<z<6$, with a very limited number of catastrophic
errors. More quantitatively, we plot the distribution of the absolute
scatter $\Delta z = (z_{spec}-z_{phot})$ in the inset: the central
part of the distribution is perfectly represented by a Gaussian
distribution (smooth red curve) with a standard deviation
$\sigma=0.06$ and a small number of outliers. Because of these
outliers, the distribution is not Gaussian and the usual r.m.s. is not
a good indicator of the width of the distribution.  If we instead adopt
the average absolute scatter, we find $<|\Delta z|>=0.08$ and
$<|\Delta z/(1+z)|>=0.045$: these values are among the lowest ever
obtained with the photometric redshift technique in the redshift
interval $0<z<6$.  Similarly, the lower panel of Fig.\ref{photoz}
shows the relative scatter $(z_{spec}-z_{phot})/(1+z_{spec})$,
restricted to the $z<2$ range and discarding the most discrepant
objects, which has a nearly homogeneous r.m.s. of 0.03. As far as
we know, this is the highest precision ever obtained on faint samples
spanning such a wide redshift range, although it is still lower than the
average SDSS value. Several elements concur to obtain this
improvement with respect to similar samples (e.g. Fontana et al 2003,
\cite{mobasher04,vanzellaNNz}): the larger number of bands, the
sophisticated technique adopted for photometry, and the use of PEGASE
2.0 models, which provided slightly better results than other synthetic
models or observed templates. We plan to discuss these effects better
in a future paper.

\begin{figure}
\includegraphics[width=9cm]{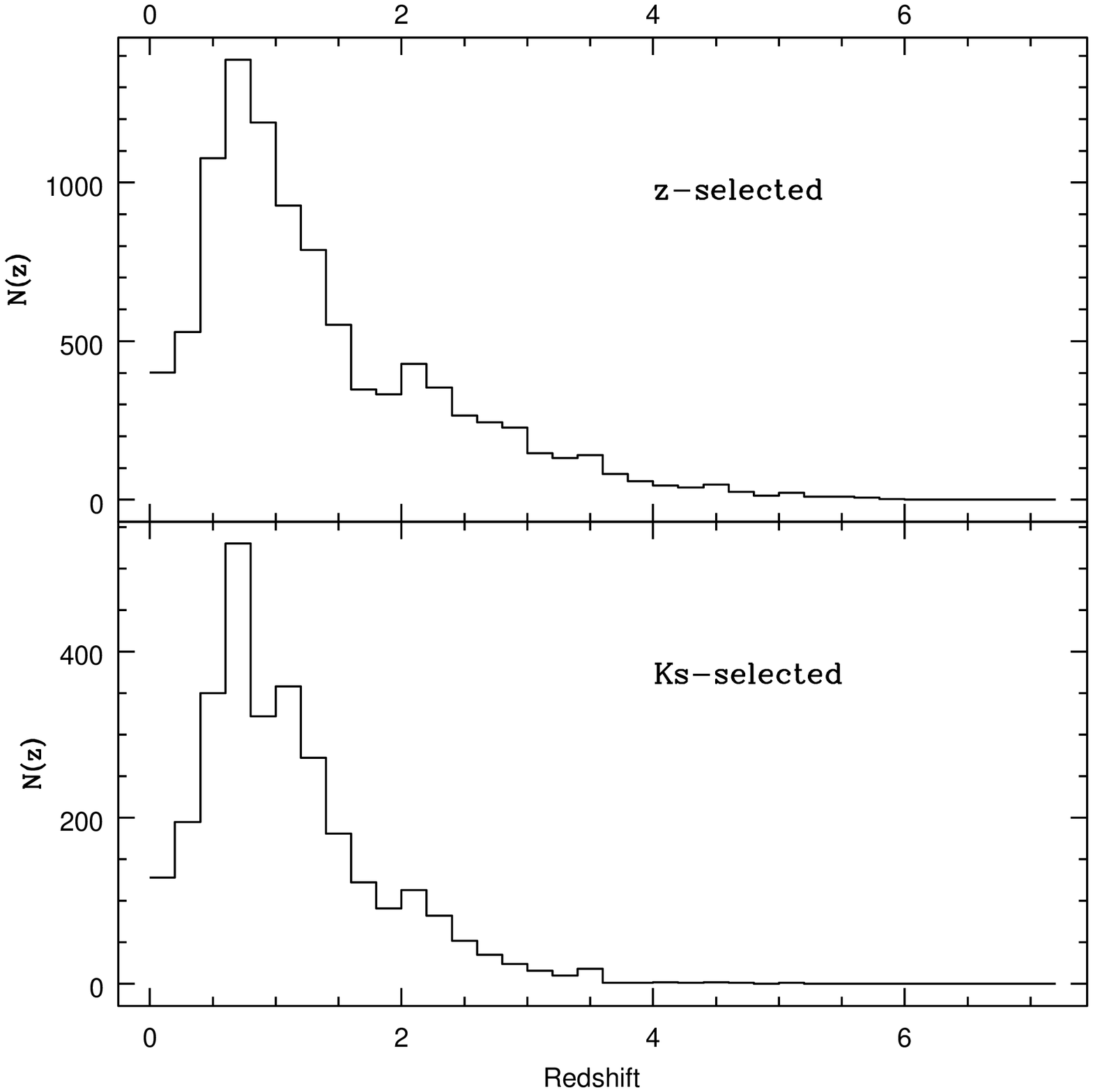}
\caption{ {\it Upper panel}: redshift distribution of 9862 galaxies in the
z--selected sample.
The typical magnitude limit is $z\simeq26$. {\it Lower panel}: redshift
distribution of 2931 galaxies in the Ks--selected sample. The typical
magnitude limit is $Ks\simeq23.8$.}
\label{nz}
\end{figure}

The resulting redshift distributions are shown in Fig.\ref{nz}, both
for the 9862 galaxies of the z--selected sample and for the 2931
galaxies of the Ks--selected one. They were drawn by adopting the
photometric redshift for all galaxies except those with a secure
spectroscopic one.  For the reasons described in Sect. 4, the two
samples do not have a unique magnitude limit, but for most of the sample
the typical magnitude limits are $Ks\leq23.8$ and $z\leq26$. The two
distributions show both a marked peak at $z\simeq 0.7$, which is a well
known feature of this field, and a marked decrease at $z>1.5$.

\section{Summary}

In this paper, we have described in detail the procedures adopted to
obtain a multiwavelength catalog of the large and deep areas in the
GOODS Southern Field covered by the deep near--IR observations
obtained with the ESO VLT. The catalog, named GOODS-MUSIC (MUlti-wavelength
Southern Infra-red Catalog), is made publicly available and
we plan to use it in several future scientific analyses. The main
features of this work are the following:

$\bullet$ The catalog is based entirely on public data: we included
the $F435W$, $F606W$, $F775W$ and $F850LP$ ACS images in our analysis,
along with the $JHKs$ VLT data, the Spitzer data provided by the IRAC
instrument (3.6, 4.5, 5.8, and 8.0 $\mu m$), and publicly available
U--band data from the 2.2ESO ($U_{35}$ and $U_{38}$) and VLT-VIMOS
($U_{VIMOS}$).

$\bullet$ The complications arising from the inhomogeneous coverage and depth
of the available images were addressed with great care. The first
is that a unique magnitude limit cannot be defined in a given band. For
this purpose, we included a careful estimate of the magnitude
limits across the detection images ($z$ and $Ks$) to make the
extraction of statistically well-defined samples possible. In practice, both 
the $z$ and the $Ks$--selected catalogs were split into several
sub-catalogs with a well-defined magnitude limit, using the bins
listed in Tab.\ref{maglimtab}. These catalogs should be used as independent
catalogs in all those cases where volume--sensitive statistics
have to be applied, as in the case of luminosity densities or
luminosity functions. Analyses that are sensitive to the position
information (such as clustering) instead requires the full use of the 
positional information.

$\bullet$ The object detection was first done in the ACS $z$
(i.e. $F850LP$) image, using a customised version of SExtractor, which
we developed to cope with the large dynamical and morphological
range of the ACS mosaics. In addition to this, we identified
the objects that are detected in the $Ks$ images that escaped
detection in the $z$ one. This double-pass procedure enabled us to
obtain a unique catalog from which either a $z$--selected or a
$Ks$--selected subsample can be easily extracted.

$\bullet$ A detailed set of simulations was executed to estimate
the completeness limits of the samples. Although the effects of
inhomogeneous coverage and depth mentioned above prevent a unique
threshold from being adopted, we find that the typical completeness is
at the level of $z\simeq 26$ or $Ks\simeq 23.8$.

$\bullet$ The galaxy colours in the ACS images were estimated
straightforwardly using the isophotal magnitudes provided by
SExtractor. Conversely, the colour estimate on the IR or UV images,
which have much poorer quality, was done using a specific
``PSF-matching'' software that we developed to this end and that we
named ConvPhot. It is described at length in Sect.4.  As a
result, all the 14847 galaxies (plus stars and AGNs) in the sample have
a $U_{35},U_{38},B,V,i,z,J,Ks,3.6,4.5,5.8,8.0$ coverage, 9651 also have
$U_{VIMOS}$ and 8441 have $H$.

$\bullet$ We cross-correlated our catalog with all the
spectroscopic surveys available to date, assigning a spectroscopic
redshift to more than 1000 sources in our catalog.  

$\bullet$ The final catalog
is made up of 14847 objects, with at least 72 known stars, 68
AGNs, and 928 galaxies with spectroscopic redshift (668 galaxies with
reliable redshift determination).  

$\bullet$ We applied our photometric redshift code to the 14 bands catalog.
We applied a standard $\chi^2$ technique, choosing
a set of synthetic templates drawn from the PEGASE2.0 synthesis model.
The comparison with the spectroscopic sample shows that the quality of
the resulting photometric redshifts is excellent, with an
r.m.s. scatter of only 0.06 for the redshift interval $0<z<6$.

$\bullet$ The full multicolour GOODS-MUSIC
catalog, including the redshift information (both
spectroscopic and photometric) is made publicly available together
with the software specifically designed for this purpose at the site
{\sf http://lbc.oa-roma.inaf.it/goods} or at CDS\footnote{
\em http://cdsweb.u-strasbg.fr/}. An example of the
public data can be found in Tables \ref{goodscatA}, \ref{goodscatB},
and \ref{goodscatC}.

%
\begin{table*}[t]
\caption[]{Extract of the GOODS-MUSIC web catalog: coordinates,
spectroscopy and magnitude limits.}
\label{goodscatA}
$$
\begin{tabular}{lccccccccccccc}
\hline
\hline
\noalign{\smallskip}
ID & RA & DEC & z$^{\mathrm{a}}$ & class$^{\mathrm{b}}$
& catalog$^{\mathrm{c}}$ & qz$^{\mathrm{d}}$ & zphot & POS$^{\mathrm{e}}$
& star$^{\mathrm{f}}$ & AGN$^{\mathrm{g}}$
& zlim$^{\mathrm{h}}$ & Kslim$^{\mathrm{i}}$
& S/G$^{\mathrm{j}}$ \\
 & J2000 & J2000 &  & & & & & & & & & & \\
\noalign{\smallskip}
\hline
\noalign{\smallskip}
10015 & 53.109565 & -27.788200 & 0.995 & emission & K20 & 0 & 0.995
& 1 & 0 & 0 & 28.044 & 26.982 & 0.030\\
10016 & 53.166175 & -27.787519 & 1.097 & galaxy & COMBO17 & 0 & 1.097
& 1 & 0 & 0 & 28.026 & 27.008 & 0.000\\
10017 & 53.124374 & -27.788923 & -1.00 & unknown & unknown & 99 & 0.300
& 1 & 0 & 0 & 28.036 & 26.957 & 0.010\\
10018 & 53.056438 & -27.788972 & -1.00 & unknown & unknown & 99 & 3.500
& 1 & 0 & 0 & 28.046 & 26.327 & 0.010\\
10019 & 53.046402 & -27.789001 & -1.00 & unknown & unknown & 99 & 0.280
& 1 & 0 & 0 & 28.051 & 26.136 & 0.740\\
10020 & 53.065872 & -27.787111 & 0.738 & early & K20 & 0 & 0.738
& 1 & 0 & 0 & 28.036 & 26.326 & 0.030\\
........ & ........... & .......... & ....... & ....... & ..... & ... & ......
& ... & ... & ... & ...... & ....... & .....\\
\noalign{\smallskip}
\hline
\hline
\end{tabular}
$$
\begin{list}{}{}
\item[($^{\mathrm{a}}$)] spectroscopic redshift (-1.0=not available).
\item[($^{\mathrm{b}}$)] spectroscopic class (see text).
\item[($^{\mathrm{c}}$)] reference spectroscopic catalog.
\item[($^{\mathrm{d}}$)] quality of spectroscopic redshift (0= very good,
1=good, 2=uncertain, 3=bad quality, 99=not available).
\item[($^{\mathrm{e}}$)] position flag (1=inside GOODS-MUSIC area, 0=outside).
\item[($^{\mathrm{f}}$)] star flag (1=probable star, 0=no star), on the basis
of spectroscopy, morphology, and BzK colours (see text).
\item[($^{\mathrm{g}}$)] AGN flag, based only on spectroscopy
(1=probable AGN, 0=no AGN). A galaxy should
have star flag=0 and AGN flag=0.
\item[($^{\mathrm{h}}$)] magnitude limit in the $z$ band in 1 sq. arcsec. and
at 1 $\sigma$.
\item[($^{\mathrm{i}}$)] magnitude limit in the $Ks$ band in 1 sq. arcsec. and
at 1 $\sigma$.
\item[($^{\mathrm{j}}$)] Star/Galaxy index of SExtractor.
\end{list}
\end{table*}

%
\begin{table*}[t]
\caption[]{Extract of the GOODS-MUSIC web catalog: optical photometry.}
\label{goodscatB}
$$
\begin{tabular}{lcccccccccccccc}
\hline
\hline
\noalign{\smallskip}
ID & $U_{35}$ & $U_{38}$ & $U_{VIM}$ & $B$ & $V$ & $i$ & $z$ &
$U_{35}$ & $U_{38}$ & $U_{VIM}$ & $B$ & $V$ & $i$ & $z$\\
 & mag & mag & mag & mag & mag & mag & mag &
err & err & err & err & err & err & err\\
\noalign{\smallskip}
\hline
\noalign{\smallskip}
10015 & 23.108 & 23.073 & 99.0 & 23.130 & 22.851 & 22.204 &
21.905 & 0.017 & 0.044 & 99.0 & 0.009 & 0.006 & 0.007 &
0.006\\
10016 & 23.085 & 22.954 & 99.0 & 23.029 & 22.618 & 21.938 &
21.425 & 0.031 & 0.070 & 99.0 & 0.017 & 0.010 & 0.011 &
0.008\\
10017 & 26.356 & -25.417 & 99.0 & 28.096 & 26.397 & 25.339 &
24.992 & 0.965 & 1.085 & 99.0 & 0.207 & 0.084 & 0.084 &
0.067\\
10018 & 26.806 & 24.918 & 99.0 & 27.257 & 25.828 & 25.444 &
25.523 & 0.820 & 0.477 & 99.0 & 0.274 & 0.046 & 0.062 &
0.086\\
10019 & 27.843 & -26.546 & 99.0 & 29.091 & 26.807 & 26.547 &
26.550 & 1.014 & 1.085 & 99.0 & 0.630 & 0.076 & 0.107 &
0.133\\
10020 & 24.893 & 24.945 & 99.0 & 24.254 & 22.429 & 20.959 &
20.473 & 0.135 & 0.386 & 99.0 & 0.044 & 0.007 & 0.003 &
0.002\\
........ & ....... & ....... & ....... & ....... & ....... &
....... & ....... & ....... & ....... & ....... & .......
& ....... & ....... & ....... \\
\noalign{\smallskip}
\hline
\hline
\end{tabular}
$$
Magnitudes are in the AB photometric system. Negative magnitudes indicate
upper limits, while $mag=99.0$ and $err=99.0$ indicate that the measure is
not available.
\end{table*}

%
\begin{table*}[t]
\caption[]{Extract of the GOODS-MUSIC web catalog: infrared photometry.}
\label{goodscatC}
$$
\begin{tabular}{lcccccccccccccc}
\hline
\hline
\noalign{\smallskip}
ID & $J$ & $H$ & $Ks$ & 3.6$\mu$ & 4.5$\mu$ & 5.8$\mu$ & 8.0$\mu$ &
$J$ & $H$ & $Ks$ & 3.6$\mu$ & 4.5$\mu$ & 5.8$\mu$ & 8.0$\mu$\\
 & mag & mag & mag & mag & mag & mag & mag &
err & err & err & err & err & err & err\\
\noalign{\smallskip}
\hline
\noalign{\smallskip}
10015 & 21.739 & 21.534 & 21.468 & 21.523 & 21.780 & 22.162 & 22.671 & 0.017 &
0.024 & 0.013 & 0.009 & 0.019 & 0.095 & 0.150\\
10016 & 20.971 & 20.662 & 20.235 & 20.142 & 20.328 & 20.793 & 20.835 & 0.011 &
0.013 & 0.010 & 0.009 & 0.010 & 0.032 & 0.033\\
10017 & 24.696 & 24.265 & 24.018 & 25.739 & -25.750 & -24.278 & -24.275 &
0.237 & 0.308 & 0.132 & 0.509 & 1.085 & 1.085 & 1.085\\
10018 & 25.579 & -26.234 & 25.129 & 25.086 & 25.071 & 23.713 & -23.970 &
0.429 & 1.085 & 0.527 & 0.412 & 0.691 & 0.816 & 1.085\\
10019 & -26.806 & 26.237 & 25.769 & 24.371 & 25.263 & 24.543 & 23.041 &
1.085 & 0.905 & 0.799 & 0.155 & 0.348 & 0.914 & 0.210\\
10020 & 20.000 & 19.496 & 19.218 & 19.267 & 19.722 & 20.007 & 20.813 & 0.005 &
0.004 & 0.004 & 0.003 & 0.004 & 0.017 & 0.034\\
........ & ....... & ....... & ....... & ....... & ....... &
....... & ....... & ....... & ....... & ....... & .......
& ....... & ....... & ....... \\
\noalign{\smallskip}
\hline
\hline
\end{tabular}
$$
Magnitudes are in the AB photometric system. Negative magnitudes indicate
upper limits, while $mag=99.0$ and $err=99.0$ indicate that the measure is
not available.
\end{table*}


\begin{acknowledgements}
It's a pleasure to thank the GOODS Team for providing all the imaging material
available worldwide. Observations were carried out using the Very Large
Telescope at the ESO Paranal Observatory under Programme IDs LP168.A-0485 and
ID 170.A-0788. We are grateful to the referee for an insightful report.
\end{acknowledgements}

\appendix

\section{True RMS estimation with correlation matrix}

We provide detailed calculations here concerning
the correlation of the noise and the theoretical interpretation.

Suppose we have a 1-dimensional distribution $x_i$, with $N$ random values.
This distribution has a mean $m$ and an r.m.s. $\sigma_x$. We define the
correlation between adjacent pixels of distance equal to $j$ as:

\begin{equation}
C_j = \frac{1}{N} \sum_{i=1}^N (x_i-m)(x_{i+j}-m) \ .
\end{equation}
The term $C_0$ is the variance of the distribution $\sigma^2_x$. If the
pixels are all randomly distributed and uncorrelated, the general term $C_j$
behaves as a Delta function:

\begin{equation}
C_j = \delta_j \sigma^2_x = \left\{ \begin{array}{ll}
\sigma^2_x & \textrm{if $j=0$} \\
0 & \textrm{if $j\neq 0$}\\
\end{array} \right.
\end{equation}
The terms $C_j$ should be nearly symmetric, if the number $N$ is large,
$C_{-n}\sim C_{+n}$.

Considering the case of a correlated distribution $y$ built from
this uncorrelated distribution, we have:

\begin{equation}
y_i=\frac{\sum_{h=-k}^{+k} w_h x_{i+h}}{\sum_{h=-k}^{+k} w_h} \ .
\end{equation}
If the kernel $w$ conserves the flux, then $\sum_{h=-k}^{+k} w_h =1$ and
$y_i=\sum_{h=-k}^{+k} w_h x_{i+h}$.

We do not demonstrate (trivial) that the two distribution $x_i$ and $y_i$ have
the same mean $m_x=m_y$. We consider here that the mean is null, $m=0$.

We now want to also evaluate the correlation coefficient $C_j$ for the new
distribution $y_i$:

\begin{equation}
C_j^y = \frac{1}{N} \sum_{i=1}^N y_i y_{i+j} \ ,
\end{equation}
which becomes

\begin{equation}
\frac{1}{N} \sum_{i=1}^N  \left [ \sum_{h=-k}^{+k}w_h x_{i+h} y_{i+j}
\right ] = \sum_{h=-k}^{+k}w_h \frac{1}{N} \sum_{i=1}^N x_{i+h} y_{i+j} \ .
\end{equation}
Now we can define

\begin{equation}
U_{hj}=\frac{1}{N} \sum_{i=1}^N x_{i+h} y_{i+j} \ ,
\end{equation}
which yields

\begin{equation}
U_{hj}=\frac{1}{N} \sum_{i=1}^N x_{i+h} \left [ \sum_{l=-k}^{+k}w_l
x_{i+j+l} \right ] \ ,
\end{equation}

\begin{equation}
U_{hj} = \sum_{l=-k}^{+k}w_l \left [ \frac{1}{N} \sum_{i=1}^N
x_{i+h} x_{i+j+l} \right ] \ ,
\end{equation}
or

\begin{equation}
U_{hj}=\sum_{l=-k}^{+k}w_l C_{j+l-h}^x \ .
\end{equation}

Now we can sum over the terms $C_j^y$:

\begin{equation}
\sum_{j=-m}^{+m} C_j^{y}=\sum_{j=-m}^{+m} \left [ \sum_{h=-k}^{+k}w_h
\sum_{l=-k}^{+k}w_l C_{j+l-h}^x \right ]  \ ,
\end{equation}

\begin{equation}
\sum_{j=-m}^{+m} C_j^{y}=
\sum_{h=-k}^{+k}w_h
\sum_{l=-k}^{+k}w_l \sum_{j=-m}^{+m} C_{j+l-h}^x \ .
\end{equation}
But $\sum_{j=-m}^{+m} C_{j+l-h}^x=\sigma^2_x$; in that case we have

\begin{equation}
\sum_{j=-m}^{+m} C_j^{y}=\sum_{h=-k}^{+k}w_h \sum_{l=-k}^{+k}w_l
\sigma^2_x=\sigma^2_x \sum_{h=-k}^{+k}w_h \sum_{l=-k}^{+k}w_l \ .
\end{equation}
Using the fact that the kernel conserves the flux, $\sum_{h=-k}^{+k}w_h=1$
and $\sum_{l=-k}^{+k}w_l=1$, it finally yields

\begin{equation}
\sum_{j=-m}^{+m} C_j^{y}=\sum_{j=-m}^{+m} C_j^{x}=\sigma^2_x \ .
\label{rmsapp}
\end{equation}
Hence, we have demonstrated that the sum of the correlation coefficient
$C_j$ gives the total variance of a distribution, whether correlated or not.
Indeed, since $C_0^y=\sigma^2_y$ and $C_0^y \le \sum_{j=-m}^{+m} C_j^{y}$,
we can conclude that the variance of a correlated distribution is less than
the variance of the original distribution, or true variance, and it is
an underestimated value for the r.m.s. of the data.

These results are derived for a discrete distribution, but they can be
generalised
to the continuum case with the help of integrals and Fourier Transform.
All these results derived from a 1-dimensional distribution can be easily
generalised for a 2-dimensional distribution, which in our case is a
reduced image, where the noise between adjacent pixels is correlated, due
to the algorithms used to sum the dithered exposures and the seeing.

In the $z$ band, the resulting kernel (Tab. \ref{rmsmat})
is positive in the central 3x3
pixels, and fades rapidly to zero at large distances. If one only
considers the central 3x3 pixels and normalises the peak value to
1, it results a correlation matrix as in Tab. \ref{rmsmat}.

\begin{table}
\caption[]{Correlation matrix of noise in the ACS $z$ band}
\begin{tabular}{cccc}
\hline
\hline
Column & -1 & 0 & +1 \\
\hline
+1 & 0.06411 & 0.28940 & 0.06131 \\
0  & 0.28436 & 1.00000 & 0.28436 \\
-1 & 0.06131 & 0.28940 & 0.06411 \\
\hline
\end{tabular}
\label{rmsmat}
\end{table}

The sum of all the terms gives the
ratio between the true and the measured variance ($\sigma_x^2/\sigma_y^2$)
of the image. 

\end{document}